A new method to identify the source vent location of tephra fall deposits: development and testing, and application to key Quaternary eruptions of Western North America

Qingyuan Yang*[(1)], Marcus Bursik[(1)], and E. Bruce Pitman[(2)]

[(1)] Department of Geology, University at Buffalo, Buffalo, NY 14260, USA

[(2)] Department of Materials Design and Innovation, University at Buffalo, 120 Bonner Hall, Buffalo, NY 14260, United States

*qyang5@buffalo.edu, (1)-716-604-4024

# Abstract

A new method to identify the source vent location of tephra fall deposits based on thickness or maximum clast size measurements is presented in this work. It couples a first-order gradient descent method with either one of two commonly-used semi-empirical models of tephra thickness distribution. The method is successfully applied to three tephra thickness and one maximum clast size datasets of the North Mono and Fogo A tephra deposits. Randomly selected and localized subsets of these datasets are used as input to evaluate its performance in cases of sparse observations. The results suggest that estimating the dispersal axis is a more robust way to constrain the source vent location with sparse observations. Bootstrap aggregating and examining the surface of the cost function are proposed to characterize the uncertainty of the method. Distinctions between the two adopted semi-empirical models of tephra thickness distribution are discussed. Results from applying the method to thickness datasets of the Trego Hot Springs and Rockland tephras are consistent with previous studies, which also provide new estimates on their total volume. The method is then applied to a series of well-correlated tephra sub-units preserved within the Wilson Creek Formation to estimate their vent location and total volume. The simplicity and flexibility of the method make it a potentially useful and powerful tool for the study of tephra fall deposits of different characteristics.

# 1 Introduction

Identifying the source vent location of tephra fall deposits is critical to the reconstruction of volcanic eruptions. Given the source vent location, characteristics of fall deposits at different sample sites can be integrated in a systematic way (e.g., Walker and Croasdale, 1971b; Self, 1983; Sieh and Bursik, 1986; Engwell et al., 2013; Klawonn et al., 2014), which is a necessary step towards further interpretation and quantification (e.g., Suzuki, 1983; Carey and Sparks, 1986; Burden et al., 2011; Pyle, 1989; Bursik et al., 1992b; Koyaguchi, 1994; Connor and Connor, 2006; Biass and Bonadonna, 2011).

For fall deposits near vent, the source location is commonly identified based on isopach and isopleth mapping and direct observation (e.g., Walker and Croasdale, 1971b; Sieh and Bursik, 1986; Miller, 1985; Bonadonna et al., 2002; Bursik et al., 2014). This is subjective, and may fail given insufficient sample sites. In such cases, identifying the source vent of a tephra deposit requires extra knowledge on the specific deposit and its potential vents (e.g., Green et al., 2014; Kawabata et al., 2015, 2016). However, the variable data quality of different tephra deposits makes it hard to come up with an universal solution. A generalized way of estimating eruption parameters, including the vent position, is to combine inverse methods with tephra transportation models (e.g., Connor and Connor, 2006; Volentik et al., 2010; Bonasia et al., 2010; Klawonn et al., 2012; Burden et al., 2013; Kawabata et al., 2013; White et al., 2017). Although sensitivity analyses have been done to these models (Scollo et al., 2008; Bonadonna et al., 2015; Pouget et al., 2016), their high-dimensional parameter space could still introduce uncertainties to the problem. This stresses the importance of uncertainty quantification (Biass and Bonadonna, 2011; Biasse et al., 2014; Green et al., 2016; White et al., 2017), which cannot be easily done given sparse observations.

For tephra found at distal sites, the identification of the source vent relies heavily on geochemical analyses as well as the geologic record (e.g., Wood, 1977; Turner et al., 2009; Pouget et al., 2014a; Marcaida, 2015; Sigl et al., 2015). The correlation with a certain vent is determined by intergrating different lines of evidence, such as similarity in age and composition.

Two semi-empirical models (Gonzalez-Mellado and Cruz-Reyna, 2010; Yang and Bursik, 2016) have been developed to analyze tephra thickness distributions for different purposes (e.g., Rhoades et al., 2002; Kawabata et al., 2013, 2016; Yang and Bursik, 2016). Kawabata et al. (2013, 2015) have used the semi-empirical model proposed by GonzalezMellado and Cruz-Reyna (2010) together with statistical approaches to recognize the number of lobes and identify the source of each lobe for different deposits, with prior knowledge on potential vent locations. The semi-empirical model proposed by Yang and Bursik (2016) is used for isopach mapping. These models describe the tephra thickness distribution as a function of location with respect to the source vent, and are similar to a cone that is stretched and rotated along the dispersal axis. This particular geometry is mainly caused by turbulent diffusion, wind advection, and horizontal spreading of the volcanic plume at the neutral buoyancy level (Csanady, 1973; Sparks et al., 1991; Bursik et al., 1992a; Rhoades et al., 2002; Bonadonna and Phillips, 2003; Gonzalez-Mellado and Cruz-Reyna, 2010; Costa et al., 2013). The two semi-empirical models are distinct in thinning rate with distance from the source vent.

In this study, we present a working version of a new algorithm that can be used to identify the source vent location of tephra fall deposits based on thickness or maximum clast size measurements. It couples either of the two semi-empirical models of tephra thickness distribution (Gonzalez-Mellado and Cruz-Reyna, 2010; Yang and Bursik, 2016) with a first order gradient descent method. Our working hypothesis is that this method can be used to identify the vent location of tephra deposits based on thickness or maximum clast size given sparse measurements. In the following text, we briefly introduce how this method works, and apply it to three different thickness plus one maximum clast size datasets of tephras from the North Mono eruptions (Sieh and Bursik, 1986) and the Fogo A tephra deposit (Walker and Croasdale, 1971b) to demonstrate its applicability. Randomly selected and localized subsets are used as input to evaluate its performance in cases of sparse observations.

In the discussion section, we propose that it is more robust to constrain the source vent location by estimating the dispersal axis first, given a sparse dataset. Bagging (bootstrap aggregating; Breiman, 1996) and examining the surface of the cost function are proposed to characterize the uncertainty of the result. Advantages and limitations of the two semiempirical models adopted for the method are discussed. The method has been further applied to thickness datasets of the Trego Hot Springs (THS) and Rockland tephras to test its performance in dealing with sparse measurements. We then apply the method to constrain the vent location and estimate the volume of a series of well-correlated tephra sub-units preserved within the Wilson Creek Formation (WCF) of Mono Lake, CA, USA (Lajoie, 1968).

The significance of our work is twofold. We present a method to constrain the source vent location of tephra deposits based on thickness or maximum clast size measurements. The method could be used to better constrain the volume and uncertainty in estimating the volume of tephra deposits, given sparse observations. At the same time, it could be used to estimate the dispersal direction and distance of tephra. The tephra dispersal pattern near-field indicates the general direction towards which most of the finer ash would travel, and therefore provides guidance on the overall distribution and extent of widely-dispersed tephra deposits. This can be used as an additional and independent constraint for the correlation of marker tephras, such as the ones preserved in the Great Basin (Davis, 1978; Madsen et al., 2002; Bursik et al., 2014). The method presented here is simple to implement, and does not necessarily require additional information about the deposit, which make it promising to be integrated into quantitative intelligent systems for tephra characterization and correlation (e.g., Bursik and Rogova, 2006; Rogova et al., 2007; Pouget et al., 2014b; Rogova et al., 2015; Kawabata et al., 2016).

# 2 Data

We build and test the method against thickness datasets of North Mono Beds 1 and 2 (digitized from Sieh and Bursik, 1986 and Bursik and Sieh, 2013; NMB1 and NMB2), the Fogo A tephra deposit (digitized from Walker and Croasdale, 1971b), and maximum clast size of NMB1 (Bursik, 1993). These deposits are selected since they are well-studied (e.g., Walker and Croasdale, 1971a; Sparks et al., 1992; Engwell et al., 2013), and display distinct features of tephra thickness/maximum clast size distribution (Yang and Bursik, 2016). The method is then applied to tephra deposits with sparse thickness measurements, including the THS and Rockland tephras, and several tephra sub-units within the WCF. Brief descriptions of the deposits and their datasets are given in this section. Their sample sites are shown in Fig. 1.

## 2.1 North Mono Beds 1 and 2

NMB1 and NMB2 are two tephra beds produced from the most recent eruptions from the Mono Craters, eastern central California. The eruptions took place during the fourteenth century A.D. (Sieh and Bursik, 1986). The rhyolitic deposits were erupted from vents at the northern end of the volcanic chain, and hence named North Mono by Sieh and Bursik (1986). The eruption produced eight distinct and widely dispersed air fall beds, as well as pyroclastic flow and surge deposits, and lava domes and flows. Isopach and isopleth maps of them are presented in the studies of Sieh and Bursik (1986) and Bursik (1993) respectively. The datasets used in this work are from Sieh and Bursik (1986); Bursik and Sieh (2013).

NMB1 is the basal unit of the North Mono tephra, and has a total volume of 0.042 $km^3$ . Its was erupted from the southwestern portion of North Coulee, and the vent is oriented along the arch of the Mono Craters (Sieh and Bursik, 1986; Bursik and Sieh, 1989, Fig. 1a inset figure). The deposit is characterized by normal grading, and was deposited in south-southwesterly winds (dispersal direction: ∼ 20°) throughout the course of the eruption, but the proximal material was transported more towards the northeast (Sieh and Bursik, 1986). More than a hundred thickness and forty-eight maximum clast size measurements (non-zero) without signs of reworking were made for this deposit. NMB2 lies above NMB1 with a sharp contact. Its vent is the Upper Dome of Northwest Coulee. The lower part of the deposit is rich in obsidian, and hence presents a darker color. This bed was blown towards the north-northwest (Sieh and Bursik, 1986; Bursik, 1993). However, the distal ( >∼ 30 km from the vent) portion of the deposit was dispersed towards the north-northeast (Sieh and Bursik, 1986). The deposit is subject to limited or no reworking, and was sampled at seventy-five sites (Sieh and Bursik, 1986).

## 2.2 Fogo Member A

The Fogo A Plinian eruption on São Miguel, Azores, took place about 5000 years ago (Moore, 1990). It has a total volume of 1.2 $km^3$ on land, and originated from Lagoa do Fogo (Walker and Croasdale, 1971b). Thickness and maximum pumice and lithic size of this deposit were sampled at more than 200 sites (Walker and Croasdale, 1971b). The deposit was further divided into syenite-poor lower (< 20 vol%) and syenite-rich upper divisions (> 20 vol%) by Bursik et al. (1992b) with the lower dispersed predominantly to the south, and the upper transported to the east (Bursik et al., 1992b). This feature does not affect the orientation of all isopachs in the hand-drawn isopach map (Walker and Croasdale, 1971b) and the one constructed from the Cubic B-spline interpolation method (Engwell et al., 2015). This two-lobe character is modeled as local variations using the method proposed by Yang and Bursik (2016). These suggest that the presence of two lobes is not evident based on thickness measurements. The dataset used in this work records the summed thickness of the two divisions. Sample sites marked with uncertainty or with no tephra observed are excluded, and the rest 184 observations are used in the present work. Although the Fogo A deposit has experienced erosion, it has been well-studied (Walker and Croasdale, 1971b; Bursik et al., 1992b; Engwell et al., 2013), and is a classical case for testing and validating tephra sedimentation models and tools for tephra deposit characterization (e.g., Bursik et al., 1992b; Kobs, 2009; Bonadonna and Costa, 2012; Engwell et al., 2015; Yang and Bursik, 2016).

## 2.3 THS and Rockland tephras and tephra sub-units in the Wilson Creek Formation

The THS tephra layer was first described by Davis (1978). It was erupted from Mt. Mazama 23.21 ± 0.30 ka, which can be corrected to ∼ 26 ka BP using the Kitigawa and van der Plicht calibration curve (Benson et al., 1997; King et al., 2001; Benson et al., 2003). A GISP2 model age has been estimated to be 29.9 ka (Benson et al., 2013). The Rockland tephra (Sarna-Wojcicki et al., 1985), aged 570-610 ka (Sarna-Wojcicki et al., 1985; Lanphere et al., 1999, 2004), was produced from a caldera now infilled by Brokeoff Volcano, a part of the Lassen volcanic complex (Clynne, 1984; Lanphere et al., 1999; Clynne and Muffler, 2010).

The two tephra deposits are critical markers for climate and stratigraphic reconstructions of the geologic history of the western United States (Benson et al., 1997; Lanphere et al., 1999; Sarna-Wojcicki, 2000; Benson et al., 2003; Adams, 2010), and were widely dispersed (Davis, 1978; Sarna-Wojcicki et al., 1985, 1991; Alloway et al., 1992). Estimated volumes of the THS and Rockland tephras are 15-56 $km^3$ (Pouget et al., 2014a) and 50-248 $km^3$ (Sarna-Wojcicki et al., 1985; Pouget et al., 2014b), respectively. In this work, we use the thickness datasets compiled by Pouget et al. (2014b) from previous studies (Sarna-Wojcicki et al., 1985, 1987; Negrini et al., 1988; Alloway et al., 1992; Rieck et al., 1992; Benson et al., 1997; King et al., 2001; Bowers, 2009; Adams, 2010; Benson et al., 2013) as inputs. Each dataset contains eight observations, and two measurements for the Rockland tephra may have been reworked (Pouget et al., 2014a).

The WCF is composed of lacustrine deposits formed in Pleistocene Lake Russell (the present-day Mono Lake), interbedded with tephra layers (Lajoie, 1968). Tephra layers within it were mostly produced from the Mono Craters. Little is known about their vent location and total volume (Lajoie, 1968; Bursik and Sieh, 1989; Marcaida et al., 2014). Among them, sub-unit correlation of Ashes B7, A4, A3, and A2 (from Black Point volcano) has been established based on outcrops sampled near Mono Lake (Yang et al., 2018a). Ash A4-d has been observed at six sites. Its stratigraphic features suggest that it was produced from the northern part of the Mono Craters (Yang et al., 2018a). Ash A4 is possibly correlative with the Lowder Creek ash found in the Markagunt Plateau, Utah (Madsen et al., 2002). Ashes B7-a, A3-f, and A2 have been observed at four to five sites. Since the method allows for the drawing of isopachs in an objective way (including estimating the wind direction objectively, given a potential vent location), it is applied to these datasets to estimate their volumes.

# 3 Method

The main idea upon which the present method is built is a gradient descent method used on a semi-empirical model of tephra thickness distribution. The concepts are introduced separately in this section. The algorithm is programmed in the R language (R Core Team, 2017), and the code is made public on Vhub (https://vhub.org/resources/4377 ; Yang et al., 2018b). Workflow of the method is shown in Fig. 2.

## 3.1 Semi-empirical models of tephra thickness distribution

The semi-empirical models proposed by Gonzalez-Mellado and Cruz-Reyna (2010) and Yang and Bursik (2016) are used here to characterize the thickness and maximum clast size distribution of tephra deposits. In this section we only refer to thickness in describing the method to avoid redundancy.

Given a known source vent location $(x_s, y_s)$ and downwind direction $(\varphi)$, the two semi-empirical models describe the tephra thickness distribution $(t_i)$ at any coordinates $(x_i, y_i)$ as:

$$t_i(r_i, \theta_i | x_s, y_s, \varphi) = \gamma e^{\{-\beta U r_i [1-\cos(\theta_i - \varphi)]\}} r_i^{-\alpha}$$

(1)

and

$$t_i(r_i, \theta_i | x_s, y_s, \varphi) = e^{[\beta_z + \beta_{dd} r_i \cos(\theta_i - \varphi) + \beta_r r_i]}$$

(2)

where $(r_i, \theta_i)$ is the polar coordinates with respect to the source vent $(x_s, y_s)$ for the arbitrary point $(x_i, y_i)$. Other parameters in the two equations (i.e., $\gamma$, $\beta$, $U$, $\alpha$ in Eq. 1 and $\beta_r$, $\beta_{dd}$, and $\beta_z$ in Eq. 2) are used to constrain the thinning rate and maximum thickness of the deposit. In practice, $\beta U$ in Eq. 1 can be combined to a single parameter.

The major distinction between the two semi-empirical models is that, given a fixed wind direction ($\varphi$) with respect to the the source (also fixed $\theta_i$), the thickness in Eq. 1 changes with both $e^{\{-\beta U r_i[1-\cos(\theta_i-\varphi)]\}}$ and ${r_i}^{-\alpha}$, but only decays with $e^{[\beta_{dd} r_i \cos(\theta_i-\varphi)+\beta_r r_i]}$ in Eq. 2. That is to say, Eq. 1 is the product of exponential and power-law functions of $r_i$, whereas Eq. 2 only thins exponentially. For simplicity, we call them the power-law (note the term ${r_i}^{-\alpha}$ in Eq. 1) and exponential models, respectively, in this study.

## 3.2 Gradient method

Given known vent position $(x_s, y_s)$ and wind direction $\varphi$, another three parameters are required to fully characterize both semi-empirical models. To simplify the problem, we take the logarithm of the two semi-empirical models. Then $t_i(r_i, \theta_i)$ can be written in matrix form:

$$A_P b_p = T \tag{3}$$

and

$$A_e b_e = T \tag{4}$$

where $T$ is a column vector with logarithmic thickness at sites $(x_i, y_i)$, $i = 1,2, \dots, n$, $b_p = (\ln(\gamma), -\beta U, -\alpha)^{\mathrm{T}}$, $b_e = (\beta_z,\ \beta_{dd},\ \beta_r)^{\mathrm{T}}$. $A_P$ and $A_e$ are n-by-3 matrices with the $\imath$th row being $(1, r_i[1-\cos(\theta_i-\varphi)],\ \ln(r_i))$ and $(1, r_i\cos(\theta_i-\varphi), r_i)$, respectively.

Eqs. 3 and 4 suggest that, once the vent position and wind direction are given with sufficient observations, the other three coefficients in Eqs. 1 and 2, respectively (i.e., vectors $b_p$ and $b_e$ in Eqs. 3 and 4), can be determined by matrix multiplication, under the assumption that residuals from the fitting follow a normal distribution. The cost function to be minimized is therefore written as:

$$f(x_s, y_s, \varphi) = \sum_{i=1}^{n}[\ln(\tilde{t}_i) - \ln(t_i)]^2 \quad (5)$$

where $\tilde{t}_i$ denotes the $i$th prediction. Minimizing this equation is equivalent to minimizing $||\tilde{T} \ - T||^2$ in matrix form, where $\tilde{T}$ is a column vector containing the logarithm of all $\tilde{t}_i$.

We implement a gradient descent method to estimate the vent position and wind direction, which is divided into two steps. First, we find the optimum wind direction that minimizes the cost function, given a proposed vent location $(\tilde{x}_s, \tilde{y}_s)$. This one-variable minimization problem can be solved using a standard one-dimensional gradient descent method. The vent location is then updated by searching in the (x, y)-plane to minimize the function:

$$g(x_s, y_s) = f(x_s, y_s, \varphi | \varphi = \tilde{\varphi}_{opt}) \quad (6)$$

where $\tilde{\varphi}_{opt}$ is the optimum wind direction derived from the gradient descent method, given fixed $(\tilde{x}_s, \tilde{y}_s)$. To find the true vent location, namely to minimize the value of Eq. 6, a two-dimensional gradient descent method is adopted. Starting from an initial guess of the vent position $(\tilde{x}_s^j, \tilde{y}_s^j)$, and a specified search radius $h^j$ with $j = 0$, the method compares the value of $g^j = g\left(\tilde{x}_s^j, \tilde{y}_s^j\right)$ and $g_*^j = \min(g\left(\tilde{x}_s^j \pm h^j, \tilde{y}_s^j \pm h^j\right))$. The procedure to find the vent location is as follows:

1. If $g^j < g_*^j$, then the true vent position is closer to $(\tilde{x}_s^j, \tilde{y}_s^j)$, and therefore, we keep $\left(\tilde{x}_s^{j+1}, \tilde{y}_s^{j+1}\right) = (\tilde{x}_s^j, \tilde{y}_s^j)$, and shrink the search radius $h^{j+1} = 0.7h^j$. The value 0.7 can be changed to any value from 0.5 to 1. The range has been chosen such that at the initial stage of the iteration, $h^j$ would not decrease too rapidly;

2. If $g^j > g_*^j$, then the vent position is closer to one of $(\tilde{x}_s^j \pm h^j, \tilde{y}_s^j \pm h^j)$, rather than $(\tilde{x}_s^j, \tilde{y}_s^j)$. Thus we need to update the vent position, which is described in the following text;

3. During the $j$th iteration, where $j > 1$, if the $(j - 2)$th iteration did not update $(\tilde{x}_s^{j-2}, \tilde{y}_s^{j-2})$ to a new location (i.e., $(\tilde{x}_s^{j-2}, \tilde{y}_s^{j-2}) = (\tilde{x}_s^{j-1}, \tilde{y}_s^{j-1})$), the $(j - 1)$th iteration did update the vent position, and we find that $g^j < g_*^j$, then that the true vent position is likely to be located within the circle centered at $(\tilde{x}_s^{j-2}, \tilde{y}_s^{j-2})$ with a radius of $h^{j-2}$. However, the $j$th step implies that the source vent is close to $(\tilde{x}_s^j, \tilde{y}_s^j)$. The two lines of evidence provide a better constraint on the source vent location, suggesting that the current step is close to the source vent region. It is thus necessary to approach with care, i.e., shrink the search radius at a greater rate, and then start the procedure anew. Therefore the method updates $h^{j+1} = \frac{(1-0.7)}{0.7} h^j$ , and keeps $(\tilde{x}_s^{j+1}, \tilde{y}_s^{j+1}) = (\tilde{x}_s^j, \tilde{y}_s^j)$ for the next iteration.

   Even if the true vent position is outside the circle centered at $(\tilde{x}_s^{j-2}, \tilde{y}_s^{j-2})$, for example, due to the particular distribution of the cost function in the x−y plane, the method could still guide the iteration to move $(\tilde{x}_s^j, \tilde{y}_s^j)$ towards the correct vent position at a slower rate. That is the reason why the shrinking rate $\frac{(1-0.7)}{0.7}$ is not set to a small value.

This setup allows the method to be approximate, but more efficient during the initial iterations. As $(\tilde{x}_s^j, \tilde{y}_s^j)$ approximates the true vent, the length of the search radius shrinks faster, which saves unnecessary iterations that simply shrink the search radius. The iteration will be terminated as it reaches certain thresholds on pre-specified search radius, number of iterations, or value of the cost function.

In Case 2, $\left(\tilde{x}_s^{j+1}, \tilde{y}_s^{j+1}\right)$ is updated based on the difference between $g^j$ and the two smallest values among $g\left(\tilde{x}_s^j \pm h^j, \tilde{y}_s^j \pm h^j\right)$, defined as $g_*^j$ and $g_{**}^j$:

a. If $g^j > g_*^j$ and $g^j > g_{**}^j$and $g_*^j$ and $g_{**}^j$ are produced due to the change in both x and y directions, then the quadrant which $\left(\tilde{x}_s^{j+1}, \tilde{y}_s^{j+1}\right)$ will be in with respect to $\left(\tilde{x}_s^j, \tilde{y}_s^j\right)$ is determined. For example, if $g_*^j = g(\tilde{x}_s^j + h^j, \tilde{y}_s^j)$ and $g_{**}^j = g(\tilde{x}_s^j, \tilde{y}_s^j + h^j)$, then the next step will move to the first quadrant with respect to $\left(\tilde{x}_s^j, \tilde{y}_s^j\right)$. The direction of the shift vector is also determined by $\Delta g_*^j = g_*^j - g^j$ and $\Delta g_{**}^j = g_{**}^j - g^j$. Following that $\left(\tilde{x}_s^{j+1}, \tilde{y}_s^{j+1}\right)$ will go to the first quadrant, $\left(\tilde{x}_s^{j+1}, \tilde{y}_s^{j+1}\right) = \left(\tilde{x}_s^j + \left|\Delta x^j\right|,\ \tilde{y}_s^j + \left|\Delta y^j\right|\right)$, and $\left(\left|\Delta x^j\right|,\ \left|\Delta y^j\right|\right) = \left(\frac{-h^j \Delta g_*^j}{\sqrt{(\Delta g_*^j)^2 + (\Delta g_{**}^j)^2}}, \frac{-h^j \Delta g_{**}^j}{\sqrt{(\Delta g_*^j)^2 + (\Delta g_{**}^j)^2}}\right)$.

b. If $g^j > g_*^j$ and $g^j > g_{**}^j$ and they are produced due to the change only in x or y direction (ridge), or $g^j > g_*^j$ and $g^j < g_{**}^j$, then we move the point towards the direction that is consistent with $g_*^j$, and fix the other coordinate. For example, if $g_*^j = g(\tilde{x}_s^j + h^j, \tilde{y}_s^j)$, the next step will be moved towards the positive direction of the x-axis, and the absolute value of the shift vector for the next step is $\left(\left|\Delta x^j\right|,\ \left|\Delta y^j\right|\right) = (h^j, 0)$.

Note that an explicit formulation of the above arguments is not given, because the method considers the comparison with $g\left(\tilde{x}_s^j \pm h^j, \tilde{y}_s^j \pm h^j\right)$ instead of approximating the gradient of the cost function. This treatment is intentionally designed to reduce the chance of falling into local minima. It is worth noting that during the implementation of the method, unphysical predictions, namely thickness distributions that thicken with distance, might occur. They can be easily recognized by examining values of the fitted coefficients: the results are unphysical if the predicted $\beta U < 0$ or $\alpha < 0$ for the power-law model (Eq. 1), or the predicted $(\beta_{dd} + \beta_r) > 0$ for the exponential model (Eq. 2). In our experiments, such unphysical predictions only occur given sparse (< 10 observations) or highly localized sample points.

### 3.3 Application to simulated data

We use points generated by the exponential model to test the method. The corresponding isopach map, source vent location, sampled points, traces of the iterative runs, and the value of the cost function (Eq. 6) for every cell within a specified grid are shown in Fig. 3. The results show that in most cases, this method is able to find the source vent location accurately given limited amount of error-free data, and that incorrect predictions could occur if the starting point for the iteration is too close to a local minimum (as illustrated in the lowerleft of Fig. 3a).

## 4 Results

In this section we present results from applying the method to datasets introduced above. First, subsets of the original datasets are used as input to demonstrate its applicability. Then more subsets with fewer input points are applied to the method to evaluate its performance. For the NMB1 and NMB2 datasets, localized observations are used as input to see if the layout of sample sites could affect the prediction.

The two semi-empirical models are considered equally important, and therefore, in each case, results from both models are presented unless specified. The starting point for the method in each application is a random point within the extent of the Mono Craters, or near the thickest sample point (within 250 m). As our results are always presented as a series of estimates, the randomness can be neglected.

## 4.1 Method illustration

Different number of data points (NMB1 and NMB2: 60, 40, and 20; Fogo A deposit: 150, 100, 50, 30) are randomly drawn from the complete thickness datasets for 25 times each, and applied to the method. The same procedure was done for the maximum clast size data of NMB1 except with 30 input points drawn from the complete dataset for 10 times due to fewer total observations. Since this section is aimed at demonstrating the applicability of the method, these subjective choices would not affect the motifs herein.

For the NMB1 (Fig. 4), when the power-law model is used, most of the predictions are located at or very close to the true vent, the North Coulee, regardless of the size of the input dataset. There are four exceptions which lie on the upwind part of the dispersal axis. With the exponential model, the predictions form a linear pattern, which is collinear to the upwind portion of the dispersal axis. Predicted source vents derived from the maximum clast size data behave in similar fashions respectively. When the power-law model is used, all ten predictions are located at the true vent location. While seven out of ten predicted vent locations (the other three are outside the extent of Fig. 4c) from the exponential model are again located in the upwind portion of the dispersal axis.

The predicted wind direction for the NMB1 with thickness data (Fig. 4d and e) is consistent with previous study (Sieh and Bursik, 1986), and the difference between results from the two models is negligible. Bimodal distribution in predicted wind direction can be observed when the maximum clast size data is used (Fig. 4f). The two modes correspond to the dispersal direction for the proximal (northeast to east-northeast) and distal (north-northeast, representing the general dispersal pattern) deposits. This feature becomes more distinct here because fewer observations are made for maximum clast size at distal sites.

Predicted vent locations for NMB2 (Fig. 5) using the power-law model is concentrated at the eastern half of the true vent, the Upper Dome. With 60 and 40 input points, forty-three out of fifty observations are located at or near the real vent. With 20 points, predictions becomes more scattered, and twelve out of twenty-five of them are within or near the contour of the vent. Similar results are obtained when the exponential model is used. The main difference is that the results from the exponential model are concentrated at the western half of the Upper Dome, which is more accurate (Sieh and Bursik, 1986). Predicted wind directions (Fig. 5c and d) between the two models are in general consistent with the main dispersal direction of NMB2, but they display a slight difference (~ 15°).

For the Fogo A deposit, using the power-law model, most of the predictions are at the southern rim of the Lagoa do Fogo caldera lake (Fig. 6a). With decreased input dataset size, predictions become more scattered, but most (> 95%) of the predictions are within 2.5 km from the vent. The fact that predicted vents are concentrated at the southern part of the true vent is due to inevitable errors introduced during map digitization. As an example, Point (a) in Fig. 6a is sampled on the island in the north of it. With the exponential model, similar results are obtained (Fig. 6b). The main difference is that predicted vents become more scattered in the longitudinal direction (increased variance in coordinates) given smaller input dataset. In terms of predictions in dispersal direction (Fig. 6c), the two semi-empirical models give similar results, pointing towards southeast (with a difference of ~ 20°), but the variance is greater when the exponential model is used. The predicted wind directions shown in Fig. 6c and d are similar to the overall dispersal direction inferred from previous studies (Walker and Croasdale, 1971b; Engwell et al., 2015).

## 4.2 Results from random subsetting

More experiments are performed in a similar way, but with fewer input data points. We obtain a thousand randomly selected subsets with 30, 20, and 10 input points respectively for each dataset. They are applied to the method, and the predicted vents are plotted as heat maps (Figs. 7-11). Since it has been shown in Figs. 4-6 that the power-law model outperforms the exponential model when applied to the NMB1 and Fogo A datasets, the latter is only applied to NMB2.

These results, except for the case of NMB2 with the power-law model, all suggest that with 30 and 20 input points, predictions from the method are highly accurate, and are mostly concentrated at or near the correct vent location. With 15 or 10 input points, more extreme outlier predictions start to occur, but the mode is still located at the correct vent location. Even though, in a more generalized perspective, predictions derived from limited input points (15 and 10) still provide useful information about the vent location. For example, distributions shown in Figs. 7-10 imply that vents of NMB1 and NMB2 are located at the northern half of the Mono Craters.

We summarize the results (Fig. 12a and b) by counting the number of predictions that are within 3000 and 2000 m from the center of the vent in each case. In general, as shown in Figs. 7-11, more than 85% (with the exception of the NMB1 maximum clast size dataset with accuracy of 69.4%) and 75% of the total predictions are within 3000 m from the true vent position given 30 and 20 input points, respectively. More than 67% and 57% of the total predictions are within 2000 m from the true vent position given 30 and 20 input points. The accuracy gradually decreases as fewer observations are used as input.

For the NMB1 (thickness and maximum clast size; power-law model) and NMB2 (exponential model), predictions are highly accurate. Even with 10 input points, more than 78% and 65% of the total predictions are still within 3000 and 2000 m from the true vent.

When the maximum clast size data is used as input, accuracy of the method is not significantly affected by the size of input, and the accuracy interestingly increases from using 15 to 10 points as input. This is because the dataset displays two very distinct thinning patterns (Fig. 1e, note the rapid and mild thinning for proximal and distal deposits respectively). If relatively smaller amount of sample points (< 20 in total) from both patterns are used as input, the method is more likely to have inaccurate results. Reducing the number of input points decreases the chance that samples from the two patterns are selected in one subset. The increase in accuracy is also explained by the fact that sample sites for the maximum clast size dataset are in general closer to the true vent location.

When the power-law model is applied to NMB2, its accuracy drops abruptly with 15 input points, as shown in Fig. 9, which is due to the incompatibility between the nature of the deposit and features of the power-law model.

For the Fogo A deposit, using the power-law model, accuracy of the results decreases accordingly with the number of input points. With 30 input points, the accuracies are ~ 91.0% (within 3000 m from the vent) and 70.4% (within 2000 m from the vent). These value drop to 57.6% and 36.3% with 10 input points. When summarizing the prediction accuracy for the Fogo A deposit, we use the center of the Lagoa do Fogo Lake as the vent location for consistency, because it is hard to objectively characterize the uncertainty from map digitization. Distributions of predicted vents in Fig. 11 shows that the accuracy shall be significantly improved if the digitization error could be corrected. The complex thickness distribution and local variations of the deposit also contribute to the variability of the predictions.

## 4.3 Results from localized subsets

Localized subsets of NMB1 and NMB2 are further applied to the method. Dataset of the Fogo A deposit is not used here due to local variations at greater scale. Since the layout of sample sites for the two deposits is oriented in the north-south direction, we take subsets of 30 points by a moving window along the longitudinal direction. For the first subsetting, we take the 30 southernmost sample points as input, and in the next subsetting, we exclude the southermost sample point, and include the 31st southernmost one in the input dataset.

We apply both semi-empirical models to the thickness and maximum clast size datasets of NMB1 and NMB2. The results (Figs. 13-15) are presented in the following way: the mean latitude of each subset is plotted in the blue inset box. The corresponding predicted vents are shown in the map with consistent color. Unphysical results are plotted as crosses both in the map (some are outside the current extent) and in the blue inset box. At the same time, the mean latitude of each subset is plotted against the corresponding predicted wind direction with consistent color (unphysical models are also labeled as crosses).

Results for NMB1 derived from thickness data show that with input points proximal to the source vent, predictions are highly accurate regardless of which semi-empirical model is used (Fig. 13). As the moving window moves northwards, predicted vent locations move towards the north-northeast slightly, which is collinear with the dispersal axis. There is a series of subsets with mean latitude below 4200000 that tends to produce unphysical results (red box in Fig. 13a and b), regardless of which semi-empirical model is used. This range of latitude corresponds to the area where the dispersal direction changes from northeast to north-northeast. The change in dispersal direction is more clearly reflected in Fig. 13c. For subsets farther north, the results are not shown as they become unstable and sensitive to the specified initial search location, and tend to produce unphysical predictions, or fall into local minima.

Results (Fig. 14) derived from the maximum clast size data of NMB1 are in general consistent with the results calculated from thickness measurements but with a few slight differences. Using both semi-empirical models with the maximum clast data, predicted vent locations do not align along the dispersal axis, but most of them are close to the true vent. The predicted vent locations do not change systematically with the mean latitude of input subsets. Subsets near the area where the dispersal direction changes also tend to produce unphysical results, but are located within a narrower latitude range (red box in Fig. 14a and b) compared with results in Fig. 13. For very proximal subsets, the predicted dispersal directions calculated from the power-law model are pointing towards south-southeast. This is perhaps related to the geometry of the vent that is oriented along this direction (Bursik, 1993, Fig. 1a).

For NMB2, most predictions (Fig. 15a and b) from using both models are located at or near the Upper Dome, and are not strongly affected by the mean latitude of input sample points. In general, the exponential model outperforms the power-law model. There are a few predictions computed from the power-law model located near the southern shoreline of Mono Lake, which corresponds to the minor mode in Fig. 9. With the exponential model, distal subsets lead to predictions in the west of Upper Dome, which is perhaps related to the change in dispersal direction from north-northwest to northnortheast for distal portion of NMB2. Predicted dispersal directions of NMB2 (Fig. 15c) are relatively stable with small fluctuations when subsets near vent are used (with five and two outlier predictions for the power-law and exponential models, respectively). It can be seen in Fig. 15c that there is a small but systematic difference between the predicted dispersal directions using the two semi-empirical models when subsets whose mean latitude is above ~ 4205000 are used. Based on our moving-window sampling scheme, this difference is introduced by the inclusion of a single measurement and exclusion of another. This implies that the power-law model is more sensitive to measurement error.

# 5 Discussion

Key findings from our experiments include:

1. This method is able to estimate the vent location of tephra fall deposits based on thickness and maximum clast size measurements. The results are highly accurate with 20 or more input points. Even with 15 or 10 input points, implementation of the method could still provide useful information to constrain the source vent location.
2. Performance of the method is affected by the specific semi-empirical model being used and characteristics of the analyzed tephra deposit. The power-law model outperforms the exponential model in cases of the NMB1 and the Fogo A deposit, and the exponential model leads to more accurate and stable results for NMB2. When the exponential model is applied to NMB1, estimated vent locations form a linear pattern that is collinear with its dispersal axis.
3. This method is able to detect the local change in dispersal direction for tephra deposits. By applying localized data to the method, the case with NMB2 (Fig. 15) shows that the exclusion or inclusion of a single measurement could greatly affect the resultant estimate if the power-law model is used.

Certain characteristics of a tephra deposit could affect the performance of the method. In Fig. 1e, thickness and maximum clast size measurement of the deposits under log-scale are plotted against the distance to their respective vents. Thickness and maximum clast size measurements of NMB1 are characterized by a rapid change in decay rate at $\sim 10 - 17$ km from the vent. This rapid change can be better fitted to the power-law model. As it is the product of power-law and exponential functions, it could thin rapidly at area proximal to the vent, and keep a slow thinning rate in distal region. This feature makes it flexible, and leads to the highly accurate estimates for NMB1.

NMB2 thins at a more stable rate. With the consistency between the deposit and the thinning pattern assumed by the exponential model, the results are more accurate compared to the ones from the power-law model.

Although Fig. 1e displays a generally stable thinning pattern with distance for the Fogo A deposit, the complexity in its thickness distribution cannot be reflected from this 2D relationship (Engwell et al., 2013; Yang and Bursik, 2016). As the power-law model is more flexible, it can be fitted better to the thickness observations, and therefore leads to more accurate predictions.

## 5.1 Constraining the vent position by estimating the dispersal axis

It is still unclear why the predicted vent positions for NMB1 form a linear pattern that is collinear with the dispersal axis when the exponential model is used. To answer this question and have a better understanding on the surface of the cost function (Eq. 6), we plot the value of Eq. 6 for every cell within the grids in Fig. 16 using the complete dataset as input for each case. Predictions at the global and local (if present) minimum and the corresponding dispersal axes are also plotted.

For the power-law model, local minimum occurs when thickness and maximum clast size datasets of NMB1 are applied. Surfaces of the cost function are irregular, and are not symmetric with respect to the dispersal axis. Interestingly, the dispersal axes corresponding to the the local minimum (south-southwest of North Coulee in Fig. 16a and c) are consistent with the true dispersal axis of the deposit using both datasets. The dispersal axis associated with the global minimum derived from the maximum clast size data is oriented towards east-northeast, which reflects the dispersal pattern of proximal deposit as previously seen. This is because most maximum clast size data were measured at sites proximal to the vent.

When the exponential model is applied to the NMB1 datasets, no local minimum occurs (Fig. b and d). Surfaces of the cost function are symmetric with respect to the predicted dispersal axes. The predicted vents are close to the local minimum calculated from the power-law model, and the resultant dispersal axes are highly consistent with the true dispersal pattern of the deposit.

Results of NMB2 and the Fogo A deposit behave similarly as in previous experiments for each semi-empirical model, and local minimum does not take place (Figs. 16e-f). For NMB2, again surface of the cost function is irregular when the power-law model is applied.

The above results show that with the power-law model, predicted vent positions could also lie near the upwind portion of the dispersal axis. A conceptual model (Fig. 17) is drawn to explain this. It represents the cross-section view of a tephra deposit along the dispersal axis. Its thickness distribution (Fig. 17a) suggests that the deposit was affected by wind during sedimentation. If fewer measurements are made at proximal sites, it is more likely that they are treated as errors instead of systematic variations. This could potentially cause the gradient-descent method adopted for this method to keep moving against the dispersal direction. For the exponential model, its exponential-thinning assumption cannot fit well to the rapid change in decay rate with distance to the vent, and causes the predicted vent to be located in the upwind area (Fig. 17c). Given the flexibility of the power-law model, the same scenario is less likely to occur. However, if the initial search location is at the upwind area and limited observations are made at proximal sites, local minimum could still occur (Fig. 17b). If little is known about a deposit, regardless of which method is used, the estimated vent locations could be in the upwind or downwind area (Fig. 17d and e). An example that the prediction is located along the downwind portion of the dispersal axis can be found in Fig. 13a and b.

The surface of the cost function in Fig. 16 for NMB1 and NMB2 indicates its strong dependence on dispersal direction; the value of the cost function changes faster along the direction perpendicular to the dispersal axis (Fig. 16a-f). Once the iterator moves to the area near the dispersal axis, the surface of the cost function becomes much flatter. This indicates that within such an area, the value of the cost function is less sensitive to the vent location, especially along the dispersal direction. Therefore, combined with the interpretation based on Fig. 16, this shows the consistency of the technique with the common-sense notion that the vent must lie along the dispersal axis.

In greater detail, ideal, physics-based models require that the vent lie along the dispersal axis. Entrainment of horizontal momentum by a volcanic plume can lead to plume bending (Bursik, 2001), or cause the development of a downwind propagating gravity current at the neutral buoyancy level (Bursik et al., 1992a). Ultimately, the horizontal transport of volcanic ash in the atmosphere is dominated by wind advection and turbulent diffusion (e.g. Suzuki, 1983; Bonadonna et al., 2005; Schwaiger et al., 2012). The overall axisymmetric (with respect to the centerline of the eruption column) spreading of tephra when there is no wind, as a result of vertical plume rise, spreading as an axisymmetric gravity current, and turbulent diffusion, becomes less apparent in the presence of strong wind or weak plume. Ideal plume spread in the absence of wind leads to axisymmetric tephra thickness distributions on the ground. In the presence of wind, the solution to an advection-diffusion equation in 2D with a continuous point source is asymmetric up- and downwind with respect to the wind direction (e.g., Csanady, 1973; Suzuki, 1983), but has planar symmetry along the wind axis. The differences between this pattern and the output of tephra transport and sedimentation models (Suzuki, 1983; Bursik et al., 1992a; Bonadonna et al., 2005; Schwaiger et al., 2012; Klawonn et al., 2012) are that the tephra is emitted from a vertical line source, and then falls at terminal velocity in the z-direction. The resulting planar symmetry in the deposit is independent of eruption parameters such as column height and total volume, and only relies on the assumption that the prevailing wind direction is constant.

On the other hand, to estimate the coordinates of the source vent, it is crucial to understand how a tephra deposit thins at places proximal to the vent, which is not always possible. The complex dynamics of volcanic plume near vent (Bursik et al., 1992b; Ernst et al., 1996) adds additional uncertainties to the thinning rate that cannot be easily characterized using simplified semi-empirical models.

In light of the above considerations, and given limited data, the most appropriate and robust way to constrain the vent location of a tephra fall deposit is to begin by estimating the dispersal axis, instead of the vent position. This is particularly true for deposits in which sedimentation was strongly affected by wind.

To demonstrate the procedure, the dispersal axes computed from the 1000 runs for NMB1 and NMB2 (results shown in Figs. 7, 9, and 10) are turned from vectors to grids. For one dispersal axis, its pixelated counterpart is a grid whose cell has the value one if it intersects the dispersal axis, and is zero otherwise. As long as the grid for each dispersal axis has the same extent and cell size, the results can be plotted cumulatively by summing up the pixel values. The results are shown in Fig. 18. The maximum clast size measurements of NMB1 contain too many proximal observations, and the Fogo A deposit was not affected by strong wind during deposition. They are not used as examples here.

For NMB1, distributions derived from the power-law and exponential models shown in Fig. 18 are highly similar and well-constrained, given 30 and 20 input points. With 15 and 10 sample points, there are more dispersal axes oriented towards the east-southeast when the power-law model is used. For the exponential model, fewer outliers show up, and most of the dispersal axes intersect or are close to North Coulee with consistent predicted dispersal directions.

Similar results are obtained for the NMB2 dataset. The dispersal axes from using the two models are restricted within a narrow range using 30 and 20 input points. The set of results derived from the exponential model is more accurate as most of these axes intersect the western half of Upper Dome. With 15 and 10 input points, more incorrect predictions, as indicated by the wrong predicted dispersal direction, start to occur when the power-law model is used. The variability of predicted dispersal directions also increases with the exponential model, but most of them are still consistent with the dispersal pattern of the deposit.

These results suggest that with sparse data, the exponential model is more robust in estimating the dispersal axis. The ratios of predicted dispersal axes from using the exponential model that are within 3000, 2000, and 1000 m from the center of vent are summarized in Fig. 12c-e. The ones for the power-law model are not shown, because some of the predicted dispersal axes are close to the true source vent location, but have incorrect predicted dispersal directions. This argument does not apply to results from the exponential model because the corresponding predicted dispersal axes that are close to the true vent locations are consistent with the dispersal pattern of the respective deposits. The comparison between Fig. 12a and b and Fig. 12c-e confirms the argument that it is more appropriate and robust to constrain the vent location of a tephra fall deposit by estimating the dispersal axis, instead of the vent location.

## 5.2 Propagating uncertainty through bagging

The complex eruptive history of Fogo A, local erosion, and measurement errors all contribute to the complex thickness distribution of the deposit (Engwell et al., 2013, 2015; Yang and Bursik, 2016). In this section, we couple the bagging (bootstrap aggregating; Breiman, 1996) approach with the method, and apply it to the thickness dataset of the Fogo A deposit. In this way, a series of predictions is generated, which allows us to evaluate the uncertainty in the predictions, and enables us to gain further insights into the deposit.

The main idea of bagging is to take samples from the original dataset uniformly with replacement. A series of subsets could be generated and used as input for a given method. In our case, five thousands subsets with 30 points are generated and applied to the method. The goal of this section is to characterize the uncertainty instead of estimating the source vent location, it is therefore justified that we can use the knowledge obtained from previous experiments. Given that the power-law model works better with the deposit, it is used here.

The predicted vent locations are shown in Fig. 19a. Predictions outside the box in the figure are assumed to be outliers. The experiment gleans 848 outlier predictions, and the corresponding input subsets are gathered and combined to a single dataset. The occurrence of each sample point in this combined dataset is summarized and plotted as a histogram in Fig. 19b. If each sample point has the same probability of contributing to outlier prediction, the histogram should resemble a uniform distribution, and their counts should fluctuate slightly around $138 \approx 848 * 30/184$ (the green horizontal line in Fig. 19b). However, the occurrence of outlier prediction is dependent on the inclusion of certain observations. We pick out these points by defining thresholds (red and yellow horizontal lines in Fig. 19b). Thicknesses of them and their nearby sample sites are marked in Fig. 19c.

The inclusion of yellow points in Fig. 19c tends to produce correct predictions. All but one of them are greater than 5 m thick. This is not surprising, as the source vent should be close to thicker sample sites.

For observations destabilizing the predictions (red points in Fig. 19c), some of them have significantly different thickness compared with their neighboring sites. For example,the deposit is much thinner at Points (a), (b), and (c) in Fig. 19c than measurements made at places farther from the vent. These samples represent extreme cases of local variation in the tephra thickness distribution (Yang and Bursik, 2016).

Such points provide little or no information about the general dispersal pattern or overall thickness distribution of the deposit, but potentially highlight local and unnoticed processes (e.g., erosion, topographic control, and secondary thickening) during and after the eruption.

There are also some red points that are not greatly different from their neighboring sites in thickness. We are not yet sure why these points have a higher chance of contributing to outlier predictions. It might be related to both the complex thickness distribution of the deposit and assumptions (e.g., formulation of the cost function and constant wind direction) adopted for this method. However, we do notice a positive example, which is Point (d) in Fig. 19c. Thicknesses of Point (d) and nearby sites imply that the deposit is composed of two lobes.

Given sufficient sample sites, more can be learned about a deposit, but the simplicity of semi-empirical models may not be able to capture certain distinctive features. It is necessary to recognize their existence. The coupled use of bagging and the method is an objective procedure aimed at selecting observations that may bear useful information to the interpretation of a tephra deposit. Given that the variability in thickness distribution caused by the two lobes of the Fogo A deposit stays at a local scale and that the deposit has experienced severe local erosion (Walker and Croasdale, 1971b; Bursik et al., 1992b; Engwell et al., 2013; Yang and Bursik, 2016), the selection of Points (a)-(d) in Fig. 19 is considered to be a preliminary success. Although there is not a concrete interpretation for each selected point, we argue that the coupled use of bagging and the method represents a conservative but more strict way to deal with raw thickness or maximum clast size measurements.

Compared with previous studies (Kawabata et al., 2013, 2015), our treatment assumes that outlier predictions are mostly derived from observations that were affected by unknown processes at different scales. Its mission ends when such sample points are selected, and further analysis is required to draw inferences from them. Practically, the choice on which method (methods proposed by Kawabata et al. (2013, 2015) and the one presented here) to use depends on the prior knowledge of the deposit and size and quality of the raw dataset.

## 5.3 Comparison between semi-empirical models

From a theoretical point of view, either semi-empirical method could be a more realistic representation of a tephra deposit. Based on different assumptions, analytical solutions of mass per unit area of tephra deposit on the ground indicate different thinning patterns (e.g., Suzuki, 1983; Bursik et al., 1992b; Koyaguchi and Ohno, 2001). Improved understanding of volcanic plumes and tephra particles adds new insights to its sedimentation (e.g., Carey and Sigurdsson, 1982; Bonadonna et al., 1998; Bursik, 2001). However, as an inverse problem that infers initial conditions based on limited observations, the simplicity of semi-empirical models becomes more important. The difference between the two models is analogous to the fitting schemes (Pyle, 1989; Fierstein and Nathenson, 1992; Bonadonna et al., 1998; Bonadonna and Costa, 2012) proposed for the $\log(Thickness) - \sqrt{Isopach\ area}$ plot to calculate the total volume of tephra deposits: the amount and quality of data, characteristics of the analyzed deposit and semi-empirical models, and the adopted fitting technique could all affect the credibility of the method.

From our experiments, it seems that the power-law model is usually better at predicting the vent location, given sufficient observations. With limited data, the exponential model is the more robust option. This is because the powerlaw model is more flexible, as it is the product of power-law and exponential functions. On the other hand, its flexibility reduces stability when dealing with sparse datasets.

To illustrate this behavior, we select 14 sample points from the NMB1 thickness dataset, including one observation that is much thicker and closer to the source vent (Point (a) in Fig. 20; measured thickness: 84 mm). We set the thickness at Point (a) to 30, 50, 84, and 150 mm, apply the modified datasets to the power-law and exponential models, respectively, and examine the predicted vent locations (Fig. 20).

Predicted vent locations with the power-law model (predicted dispersal direction: 2−7°) all concentrate at a location near Point (a), except when Point (a) is set to 150 mm thick, for which the corresponding predicted vent is located near North Coulee. For the exponential model, most predictions are located near the true dispersal axis with a narrow range (bounded by the two dispersal axes plotted), and the predicted dispersal directions (16−21°) are consistent with the true transport direction. This example shows concretely that given sparse data, the power-law model is sensitive to fluctuations of a single measurement. Therefore when dealing with sparse observations, it is better to concentrate on the exponential model, or to combine the power-law model with other statistical methods, or incorporate prior knowledge about the deposit (Kawabata et al., 2013, 2015, 2016; Green et al., 2016).

## 5.4 Application to sparse datasets

The above discussion justifies the use of the present method to identify or constrain the vent location of tephra deposits, and also provides examples of how to work with the method, given datasets of different size and quality. Its success in working with datasets of NMB1, NMB2, and the Fogo A tephra deposits encourages us to apply it, with the exponential model, to the THS and Rockland tephras. These are widely dispersed and significant Quaternary deposits, and their primary thickness has only been measured at eight sites each. The Albers Equal Area Conic projection system, centered at (−120° ,42.5°), is used in these two cases. We also try the same projection system centered at (−120.0°, 45.5°), which leads to consistent results.

Six sample sites of the THS tephra have been observed in northwestern Nevada, where the thickness shows great and systematic variation within a narrow swath. This suggests that transport was subject to a strong prevailing wind. Local minima tend to occur, given the paucity of sample sites, and it is necessary to examine the surface of the cost function. For the Rockland tephra, the sample sites span a broader area, and local minima do not occur in the surface of the cost function (not shown to avoid complexity). Since two of the thickness measurements might be affected by reworking for the Rockland tephra, we include or exclude them to generate two input datasets. The exponential model is used, given the limited observations in each case.

Results for the predicted vent sites and dispersal axes for the THS and Rockland tephras are shown in Fig. 21a and b, respectively. They are derived from using different initial search locations within the extents. The surface of the cost function for the THS tephra in the x-y plane is plotted in Fig. 21a. Three results for vent sites are obtained with one (green triangle in Fig. 21a) being unphysical; the others are close to Medicine Lake and Mount Shasta. Medicine Lake and Mount Mazama are near the two, respective dispersal axes. The two predictions arise from the uncertainty associated with the two sample sites in California, which could be on the same or opposite side of the dispersal axis. The fact that this deposit was affected by strong wind suggests that we should use the dispersal axes to constrain the vent location. Therefore, the results narrow down the seven potential vents to two, namely Medicine Lake and Mount Mazama. The latter is the true vent for the deposit.

Two estimated vent sites plus one unphysical prediction are obtained for the Rockland tephra. Both are near Lassen Peak (Fig. 21b). Thicknesses at the sample sites indicate that the tephra was widely dispersed downwind, but did not travel far in the upwind direction. Therefore, the source vent cannot be Mount Shasta, Medicine Lake, or the others farther north. The vent should be located near the two predicted locations shown in Fig. 21b. This inference is correct, as the Rockland tephra was produced from Brokeoff Volcano, a part of the Lassen Volcanic complex (Clynne and Muffler, 2010; Pouget et al., 2014a).

We have estimated volumes of the two deposits based on the exponential model, with the correct vent locations. We use volume estimation methods proposed by Pyle (1989); Fierstein and Nathenson (1992); Nathenson and Fierstein (2015) to further process the isopach data. Deposit volumes of the THS and Rockland tephras are estimated to be 20.8 and 326.7$km^3$, respectively. The estimate for the THS is consistent with the previous estimate (Pouget et al., 2014a), while the estimate for the Rockland tephra is beyond the range (50-248 $km^3$) found in previous studies (Sarna-Wojcicki et al., 1985; Pouget et al., 2014a). Given the methods used in the previous studies, we believe the current estimate of deposit volume represents a better, more objective, and unbiased value.

## 5.5 Broader applications

We now apply the method, coupled with the exponential model, to the thickness dataset of the WCF Ash A4-d, to constrain its vent location. In another application, the volumes of the WCF Ashes B7-a, A4-d, A3-f, and A2 are estimated, using the exponential model with four to six observations.

The measured thickness of Ash A4-d does not vary greatly from site to site, indicating that measurement uncertainty might affect the final estimate. To see if certain measurements would have a greater impact on the estimate, we use the complete dataset and its subsets, following the leave-one-out sampling scheme, as inputs to the method (Fig. 21c). All but one of the predicted vent locations are consistently located in the west of the Mono Craters, with the corresponding dispersal axes pointing towards the north-northeast. The consistent results suggest that the source vent for this subunit might be located near Dome 11. The light green prediction along the eastern shoreline of Mono Lake is derived from excluding the light green observation north of Mono Lake (Fig. 21c). This prediction comes with unphysical fitted coefficients for the semi-empirical model, and the dispersal axis is thus not plotted. The inferred vent area for Ash A4-d is consistent with a previous, very poorly constrained and non-objective interpretation based on stratigraphic characteristics (Yang et al., 2018a). To make this inference more concrete, future fieldwork should focus on the stratigraphy of A4-d at sites north of Mono Lake.

Volumes of Ashes B7-a, A4-d, A3-f, and A2 are estimated following the same procedure done for the THS and Rockland tephras. We assume different potential vent locations (Marcaida et al., 2014 and this study) for these deposits, apply the method to calculate the wind direction, and estimate the volume.

Estimated volumes shown in Table 1 suggest that given the current sample sites, the volume of these tephra layers is not significantly affected by the assumed vent location. The volume of Ash B7-a is ~0.092-0.186 $km^3$, while the others have estimated volumes below 0.05 $km^3$. Ash B7-a is thus the most voluminous single fall deposit yet known from the Mono Craters, while the range for the other deposits is consistent with previously measured volumes of Holocene fall deposits of the Mono-Inyo Craters (Table 1, Miller, 1985; Sieh and Bursik, 1986; Nawotniak and Bursik, 2010; Bursik et al., 2014).

We believe that the crude estimates on vent location and volume of the listed ash sub-units in the WCF can be used to improve our understanding on the volcanic history of the Mono Craters (Bailey, 2004; Hildreth, 2004; Marcaida, 2015; Yang et al., 2018a). They also provide new constraints on the volcanic hazard assessment of the Long Valley-Mono Craters region (Bevilacqua et al., 2017, 2018).

# 6 Conclusions

We have presented a new algorithm that can be used to estimate the source vent location of a tephra fall deposit, based on thickness or maximum clast size measurements. The method is composed of a semi-empirical model that describes the thickness or maximum clast size distribution of tephra deposits, coupled with a gradient-descent method. There are two semi-empirical models (Gonzalez-Mellado and Cruz-Reyna, 2010; Yang and Bursik, 2016) that can be used. The method is worked out and validated on datasets of North Mono Beds 1 and 2, and the Fogo A tephra deposit. The results show that the method is able to accurately and precisely predict the source vent location, given ≥ 20 sample points. With 6 to 15 input points, the method provides useful and previously unavailable constraints on the source vent location. The method performs poorly given ≤ 5 observations. With sufficient local input, the method can detect a local change in dispersal direction. The performance of the method is affected by the quality as well as the size of the input dataset, and the semi-empirical model used.

Our experiments show that, in the case of limited observations (≤ 15), estimating the dispersal axis first, instead of attempting to estimate the vent coordinates *de novo*, is a more robust way to constrain the vent location. This is because the planar symmetry of tephra thickness or maximum clast size distribution with respect to the dispersal axis, under the assumption of constant wind direction, is independent of many other eruptive and atmospheric parameters, and can be easily detected and captured by the present method. Estimating the dispersal axis first, instead of the exact coordinates of vent position, is also an appropriate choice for future studies with similar goals, and hence cannot be neglected.

Two measures are adopted to characterize the uncertainty of the method, namely bagging and examining the surface of the cost function in the x-y plane. The former stresses the effect of epistemic uncertainty due to the simplicity of the semi-empirical models, and the latter can be used to detect the occurrence of local minima.

Inter-comparison between the two semi-empirical models used in the present work suggests that it is necessary to consider or to examine characteristics of the raw measurements before choosing which model to use. The power-law model is flexible and likely to give a more accurate estimate given sufficient observations (> 20). However, for datasets with sparse (≤ 15) or unevenly-distributed sample sites, the resultant prediction from the power-law model tends to stay close to the thickest observation, whether this is in some sense near the vent or not, as a result of its flexibility and our assumption on error distribution. Under such circumstances, we recommend the use of the exponential model, owing to its stability. The method, with the exponential model, is applied to sparse thickness datasets of the Trego Hot Springs and Rockland tephras, and does well in constraining their vent location. New estimates on their volume are given following the exponential model, which yields 20.8 and 326.7 $km^3$ (~ 8 and ~ 130 $km^3$ DRE), respectively. If these estimates are correct, then the Rockland tephra is among the most voluminous fall deposits of the late Quaternary in North America, formed 30−70 kyr after the caldera-forming event from the Yellowstone Volcano which produced the Lava Creek Tuff (Lanphere et al., 2002).

The method assumes a constant wind direction, and the thickness and maximum clast size data are log-transformed (~exponential decrease in thickness and grain size) prior to the fitting process. These assumptions are parsimonious, conforming with Occam's razor, which might introduce potential uncertainties to the final estimate; however, the method is shown to be useful and efficient, and can be applied to tephra thickness or maximum clast size datasets of varying size and quality. The simplicity of the method makes it convenient and flexible to be incorporated into statistical systems and tephra databases to help identify the source vent location and correlate tephra deposits.

Application of the method to tephra sub-units within the Wilson Creek Formation shows that Ash A4-d was erupted from a vent near Dome 11 of the Mono Craters, and that at least one sub-unit of A4 was blown towards the northeast. From a transport pattern point of view, this suggests that the Lowder Creek ash could be Ash A4 (Madsen et al., 2002), as speculated by the original discoverers of that outcrop. The volume of Ash B7-a is estimated to be ∼ 0.09-0.18 $km^3$, and volumes of other analyzed sub-units are below 0.05 $km^3$. This provides new insights into the eruptive history of the Mono Craters during the late Quaternary.

## Acknowledgements

This work was supported by National Science Foundation Hazard SEES grant number 1521855 to G. Valentine, M. Bursik, E.B. Pitman and A.K. Patra and National Science Foundation Division of Mathematical Sciences grant number 1621853 to A.K. Patra, M. Bursik, and E.B. Pitman. The opinions expressed herein are those of the authors alone and do not reflect the opinion of the NSF. S. Engwell is thanked for kindly sharing the data. We thank the reviewers for commenting on an earlier version of this manuscript.

# Table and figure captions

Table 1: Estimated volumes of the THS and Rockland tephras and Ashes B7-a, A4-d, A3-f, and A2 within the WCF. Volumes of selected tephra deposits from the most recent eruptions from the Mono-Inyo Craters are also listed for reference (Miller, 1985; Sieh and Bursik, 1986; Nawotniak and Bursik, 2010; Bursik et al., 2014). Number of sample sites and corresponding predicted dispersal direction for each deposit are listed in brackets in the first and third columns, respectively. MC is short for Mono Craters.

Fig. 1: Sample sites of NMB1 (a and b for thickness and maximum clast size, respectively) and NMB2 (c), and the Fogo A deposit (d) with corresponding source vent marked as red polygon. The vent geometry of NMB1 and NMB2 with respect to the corresponding dome is also shown in the inset figure in a and c (Sieh and Bursik, 1986). e: Observed thickness (mm) and maximum clast size (mm) under log-scale are plotted against distance (m) to the source vent for each dataset.

Fig. 2: Workflow of the method. Note that numbers and letters marked in boxes correspond to method description in text.

Fig. 3: Illustration of how the gradient method works with simulated data. The exponential model (Eq. 2) is used for data simulation. The source vent location is at the origin (red triangle), and it has a wind direction of 20° with $\beta_z$, $\beta_{dd}$, and $\beta_r$ equal to 1.2, 0.01, and -0.02 respectively. The isopachs are 1.1, 0.9, 0.7, 0.5, 0.2 thick outwards. Ten (a) and twenty (b) sites (green dots) are generated and sampled as inputs. Starting from four different locations in each case, seven of the eight predictions correctly predict the source vent location, and the blue dots denote the trace of the gradient descent method. The case with local minimum starts at Point (A) in (a). Value of the cost function at each location within the plane is calculated and plotted as grids.

Fig. 4: a and b: estimated vent locations of NMB1 from using the power-law (triangle) and exponential (diamond) models respectively with 80 (red), 40 (yellow), and 20 (blue) input points randomly drawn from the original datasets. Shown are ensembles of 25 predictions, each starting at a randomly selected initial location near the Mono Craters. Three outlier predictions outside the extent of b are not shown. c: predicted vent positions of NMB1 from using the maximum clast size data with the power-law (light green triangle) and exponential (purple diamonds) models respectively with 30 input points randomly drawn for ten times from the original dataset. Three predictions from using the exponential model are outside the extent of c. d-f: predicted wind directions corresponding to a-c with consistent color.

Fig. 5: a and b: estimated vent positions of NMB2 from using the power-law (triangle) and exponential (diamond) models respectively with 60 (red), 40 (yellow), and 20 (blue) input points randomly drawn from the original thickness dataset. Shown are ensembles of 25 predictions, each starting at a randomly selected initial location near the Mono Craters. Eight and five predictions are outside the extents of a and b respectively. c and d: predicted wind directions corresponding to a and b with consistent color.

Fig. 6: a and b: estimated vent locations of the Fogo A deposit from using the power-law (triangle) and exponential (diamond) models respectively with 150 (yellow), 100 (blue), 50 (light green), and 30 (light pink) input points randomly drawn from the original thickness dataset. Shown are ensembles of 25 predictions, each starting at a randomly selected location within the sampled area. c and d: predicted wind directions corresponding to a and b with consistent color. Point (a), sampled on the island in the north of it, is marked to denote the error introduced from map digitization.

Fig. 7: Distribution of predicted vent locations of NMB1 from using the power-law model with thickness dataset. Results shown in a-d are derived from using 30, 20, 15, and 10 points as input. For each case, we randomly generate 1000 subsets to characterize the distribution of predictions. The value for each cell represents the number of predictions within it.

Fig. 8: Distribution of predicted vent locations of NMB1 from using the power-law model with maximum clast size dataset. Results shown in a-d are derived from using 30, 20, 15, and 10 points as input. For each case, we randomly generate 1000 subsets to characterize the distribution of predictions. The value for each cell represents the number of predictions within it.

Fig. 9: Distribution of predicted vent locations of NMB2 from using the power-law model with thickness dataset. Results shown in a-d are derived from using 30, 20, 15, and 10 points as input. For each case, we randomly generate 1000 subsets to characterize the distribution of predictions. The value for each cell represents the number of predictions within it.

Fig. 10: Distribution of predicted vent locations of NMB2 from using the exponential model with thickness dataset. Results shown in a-d are derived from using 30, 20, 15, and 10 points as input. For each case, we randomly generate 1000 subsets to characterize the distribution of predictions. The value for each cell represents the number of predictions within it.

Fig. 11: Distribution of predicted vent locations of Fogo A deposit from using the power-law model with thickness dataset. Results shown in a-d are derived from using 30, 20, 15, and 10 points as input. For each case, we randomly generate 1000 subsets to characterize the distribution of predictions. The value for each cell represents the number of predictions within it.

Fig. 12: a and b: ratio of predictions that are within 3 and 2 km from the center of the true vent for different datasets using different semi-empirical models. It summarizes the results shown in Figs. 7-11. c-e: ratio of predicted dispersal axes that are within 3, 2, and 1 km from the center of the correct vent for NMB1 and NMB2 (using thickness data as input). Results from using the power-law model are not shown, because some predicted dispersal axes from the power-law model are close to the true vent location, but have incorrect predicted dispersal direction.

Fig. 13: a and b: estimated vent positions of NMB1 derived from the power-law (triangle) and exponential (diamond) models using localized thickness observations (size: 30). The subsetting is done using a moving window scheme along the longitudinal direction (see text for detailed explanation). The mean latitude of each subset is plotted in the blue box. Corresponding predicted vents are plotted in the map with consistent color. Crosses in the blue box and map represent unphysical predictions. Subsets whose mean latitude is above the current extent tend to produce unphysical predictions, and thus the corresponding predictions are not shown. c: mean latitude of each subset is plotted against the corresponding predicted dispersal direction (in degree from North clockwise). The occurrence of crosses and color of the points are consistent with a and b.

Fig. 14: a and b: estimated vent positions of NMB1 derived from the power-law (triangle) and exponential (diamond) models using localized maximum clast size observations (size: 30). The subsetting is done using a moving window scheme along the longitudinal direction (see text for detailed explanation). The mean latitude of each subset is plotted in the blue box. Corresponding predicted vents are plotted in the map with consistent color. Crosses in the blue box and map represent unphysical predictions. c: mean latitude of each subset is plotted against the corresponding predicted dispersal direction (in degree from North clockwise). The occurrence of crosses and color of the points are consistent with a and b.

Fig. 15: a and b: estimated vent positions of NMB2 derived from the power-law (triangle) and exponential (diamond) models using localized thickness observations (size: 30). The subsetting is done using a moving window scheme along the longitudinal direction (see text for detailed explanation). The mean latitude of each subset is plotted in the blue box. Corresponding predicted vents are plotted in the map with consistent color. No unphysical predictions are obtained in this case. c: mean latitude of each subset is plotted against the corresponding predicted dispersal direction (in degree from North clockwise). Color of the points is consistent with a and b.

Fig. 16: Surface of the cost function (Eq. 6) and predicted vent location and dispersal axis using the complete dataset as input. a, c, e, and g correspond to results for NMB1 (thickness and maximum clast size data), NMB2, and the Fogo A deposit derived from the power-law model.

Predictions at local minimum are also displayed (as void triangle and dashed line) if present. b, d, f, and h correspond to results for NMB1 (thickness and maximum clast size), NMB2, and the Fogo A deposit derived from the exponential model (yellow diamond and solid lines). Local minimum does not take place when the exponential model is used. The results from using the power-law model are also displayed in b, d, f, and h for easier comparison.

Fig. 17: Conceptual model explaining how characteristics of a deposit and the layout of sample sites could affect the estimated results. The figures denote the cross-section vew of a tephra deposit whose deposition was affected by wind. Fewer measurement (yellow dots) are made for proximal and upwind portion of the deposit. a: the deposit is characterized by rapid thinning proximal to the vent, and thinning rate for distal deposit downwind is small. b: using power-law model, its flexibility could lead to the correct answer, but due to the lack of measurements for the proximal and upwind deposit, the rapid change in thickness might be regarded as non-systematic, which could result in the predicted vent located in the upwind area. c: the exponential model assumes a constant thinning rate under log-scale. It is less flexible compared with the power-law model. Therefore the predicted vent location would be located in the upwind area following the same argument in b. d and e: if little is known about a tephra deposit or only distal measurements are made for a deposit, the predicted vent location is hard to infer using both models. In such cases, a more robust alternative option is to estimate the dispersal axis as argued in text.

Fig. 18: Distributions of the resultant dispersal axes from the 1000 run results shown in Fig. 7, 9, and 10 for NMB1 and NMB2. Results for NMB1 with the exponential model are results not shown

in previous figures, but are derived from the same input data subsets. The value of each cell represents the number of dispersal axes that intersect it. As the size of input dataset decreases, results from the power-law model become less stable.

Fig. 19: a: predicted vent locations (blue triangles) of the Fogo A deposit from 5000 randomly-selected subsets with 30 points using the power-law model. Predictions outside the bounding box are assumed to be outliers. Input subsets leading to outlier predictions are collected and combined to a single dataset, and the occurrence (y-axis of b) of each observation within the combined dataset is plotted as a histogram in b (x-axis: index of each sample point). There are 848 outlier predictions, which suggests a total number of $25440 = 848 * 30$ points are drawn. If the likelihood of contributing to outlier is the same for each sample, the expected value in b should be $138 \approx 848 * 30/184$ (green horizontal line in b). Some points have a higher or lower chance of contributing to outlier predictions. Such points are selected based on two thresholds marked as yellow and red horizontal lines in b. Thicknesses (m) of them and their neighboring sites are marked in c.

Fig. 20: Fourteen points from NMB1 thickness dataset are selected as input to highlight the sensitivity of the power-law model to thicker measurement given sparse data. The deposit at Point (a) is 84 mm thick, and is much thicker than the rest input points (green dots). The thickness of Point (a) has been reset to 50, 84, 100, and 150 mm. Each of this value is combined with the rest input points, and the combined dataset is applied to the power-law and exponential models. Only one of the results from the power-law model (red triangle) are located near North Coulee (when the thickness of Point (a) is set to 30 mm), and the rest of them are all located near Point (a).

Predicted vent locations (yellow diamond) using the exponential model are all located near the true dipsersal axis, and the predicted dispersal directions are all within a reasonable range (16.97−20.50° from North clockwise) that is consistent with the dispersal pattern of the deposit. Two dispersal axes derived from the exponential model are shown that correspond to the predictions when Point (a) is set to 30 and 150 mm thick respectively. They represent the end members of the predicted dispersal axes. See text in the figure for more details. This highlights that the power-law model is less stable given sparse dataset using the present method, and is sensitive to the thicker measurement. This suggests that estimating the dispersal axis with the exponential model is a more robust way to characterize the vent location given limited observations.

Fig. 21: Estimated vent locations (red triangle) and dispersal axes (black arrow lines) for the THS (a) and Rockand (b) tephras and the WCF Ash A4-d (c). Thickness (cm) of the deposits at each sample site (green dot) is labeled. Unphysical predictions are marked as green triangle. In a and b, potential vents (following Pouget et al. (2014a)) for the two deposits are plotted as void triangle. They are derived from using different initial search locations within the extents of the figures. For the THS tephra, the surface of the cost function is plotted, and two predictions plus one unphysical prediction are obtained. For the Rockland tephra, two sets of results plus one unphysical prediction are derived from including and excluding two possibly reworked observations. For the WCF Ash A4-d, results from using the complete dataset (predicted vent location marked with darker red triangle) and subsets from leave-one-out sampling scheme are shown. There is one unphysical prediction that is derived from excluding the light green observation in the north of Mono Lake in the input. We draw a red circle to denote the inferred vent area for the deposit based on estimated

dispersal axes. Locations of Domes 8, 11, 25, and 31 of the Mono Craters are labeled.

Table

Table 1

| Tephra unit/sub-unit | Assumed vent Location/ Eruption name | Estimated Volume ($km^3$) |
|---|---|---|
| THS (8) | Mt Mazama | 20.8 |
| Rockland (8) | Brokeoff Volcano | 326.7 |
| B7-a (4) | MC Dome 8 | 0.186 (243.20) |
| | MC Dome 31 | 0.115 (261.52) |
| | MC Dome 25 | 0.092 (289.57) |
| A4-d (6) | MC Dome 8 | 0.041 (48.06) |
| | MC Dome 11 | 0.041 (47.60) |
| | MC Dome 31 | 0.034 (35.29) |
| | MC Dome 25 | 0.031 (26.68) |
| A3-f (5) | MC Dome 8 | 0.017 (339.90) |
| | MC Dome 11 | 0.017 (339.30) |
| | MC Dome 31 | 0.014 (341.68) |
| | MC Dome 25 | 0.015 (349.11) |
| A2 (5) | Black Point | 0.047 (297.52) |
| Orange-brown beds | South Mono | 0.0156 |
| Basal Beds | | 0.0054 |
| North Mono Bed 1 | North Mono | 0.042 |
| South Deadman 1 | 1350 A.D. Inyo eruption | 0.01 |
| South Deadman 2 | | 0.04 |
| Lower Obsidian Flow | | <0.01 |
| Upper Obsidian Flow | | 0.01-0.02 |
| Glass Creek | | 0.1 |

Figures

Figure 1

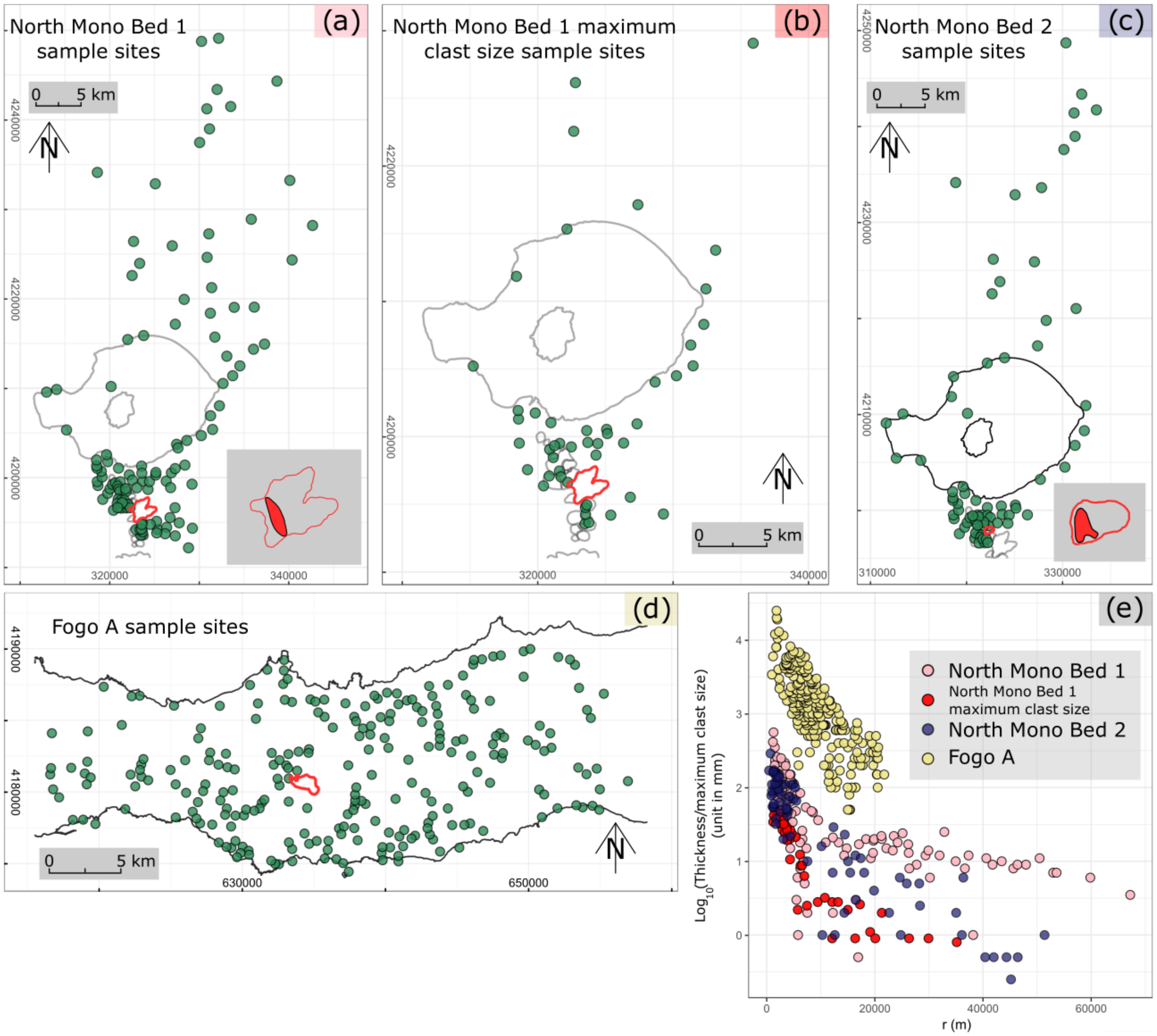

Figure 2

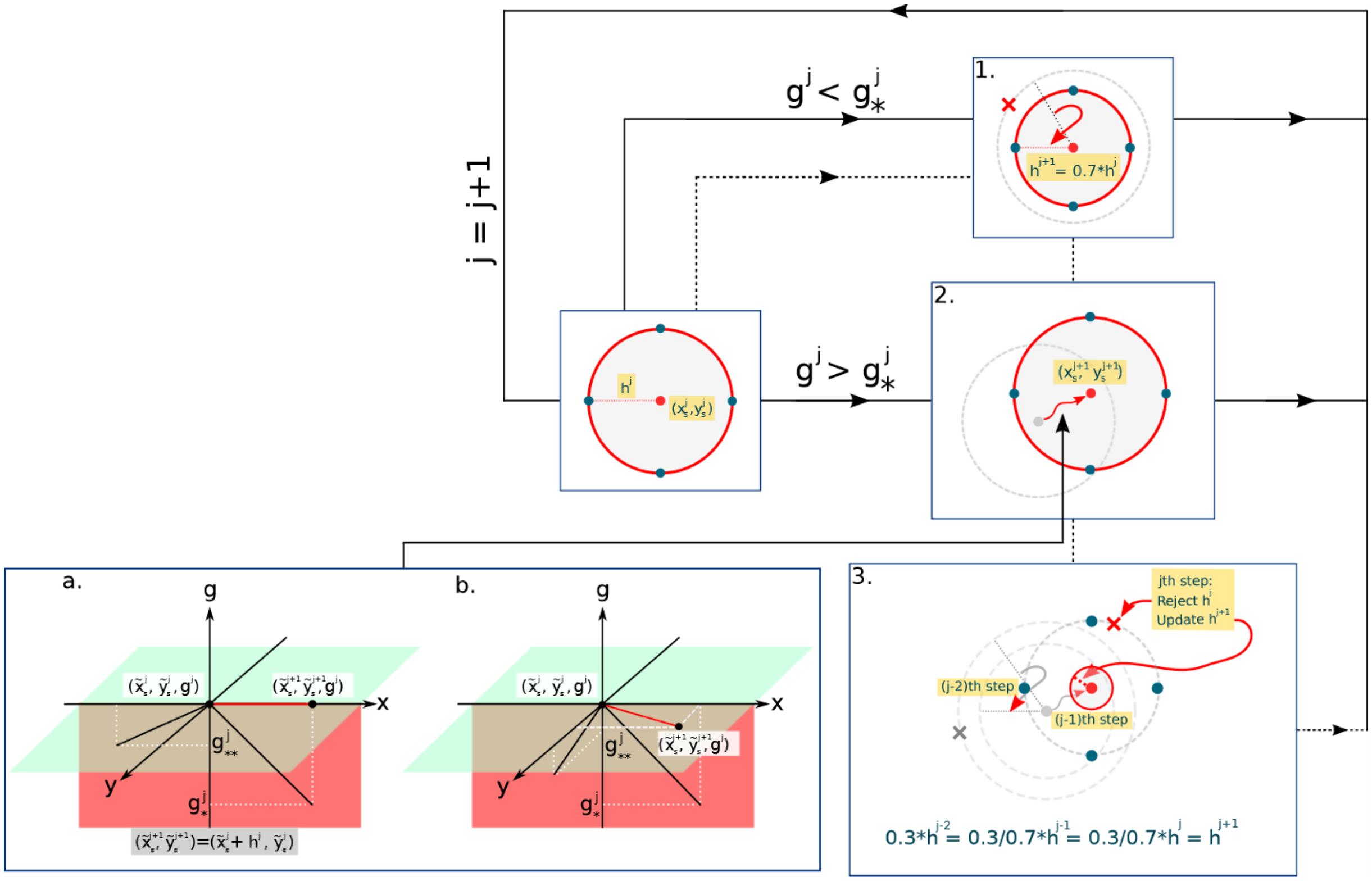

Figure 3

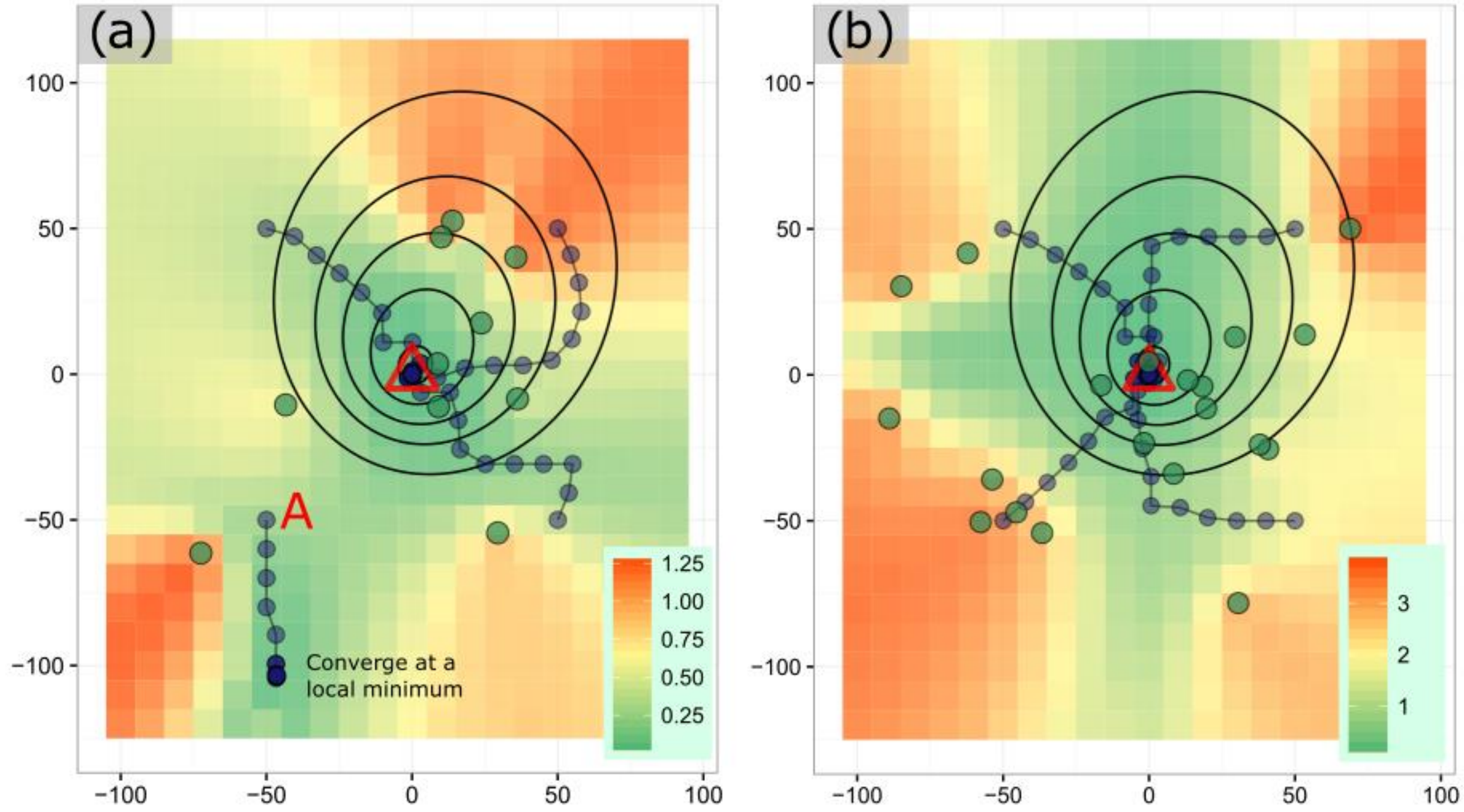

Figure 4

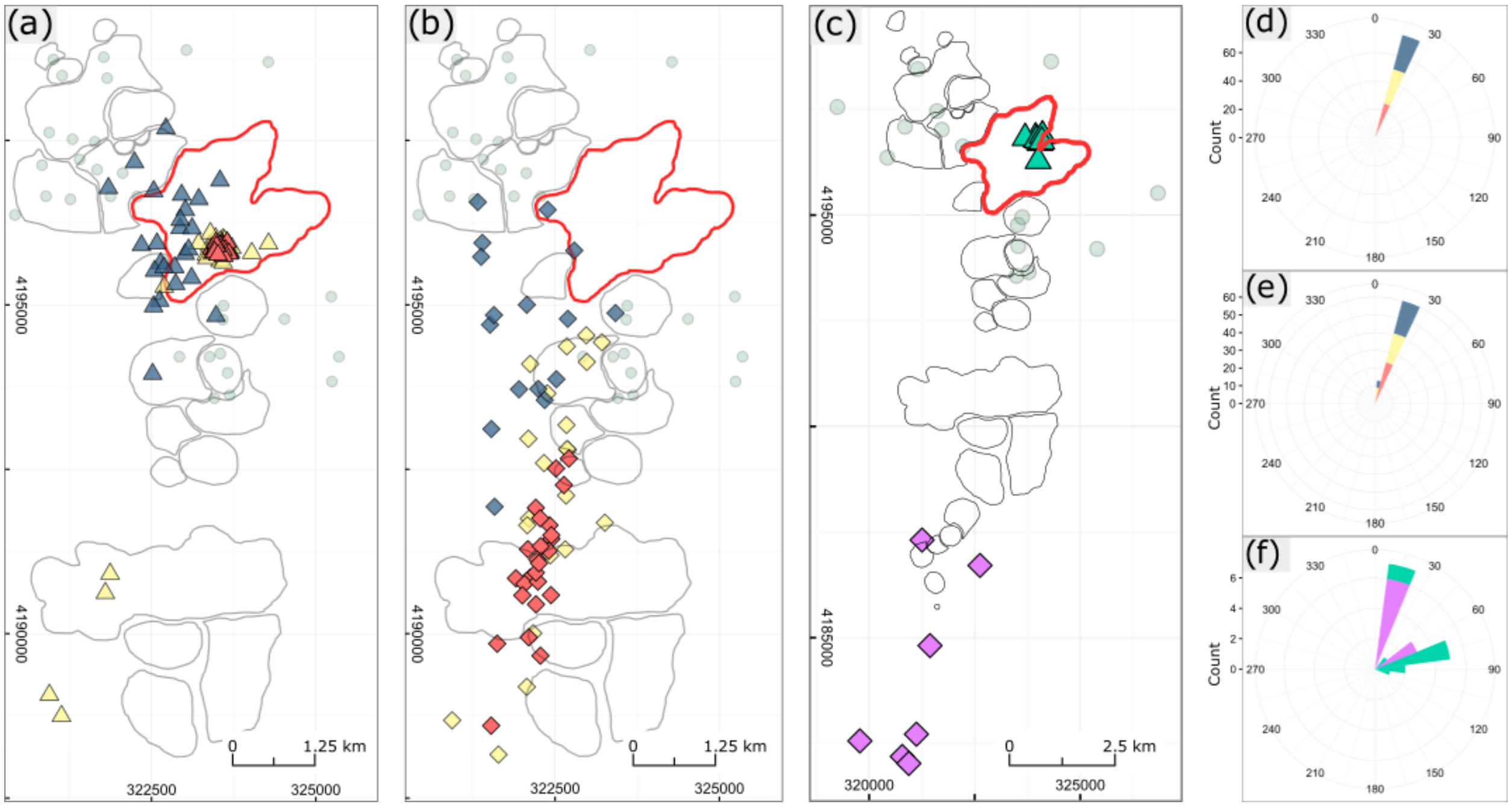

Figure 5

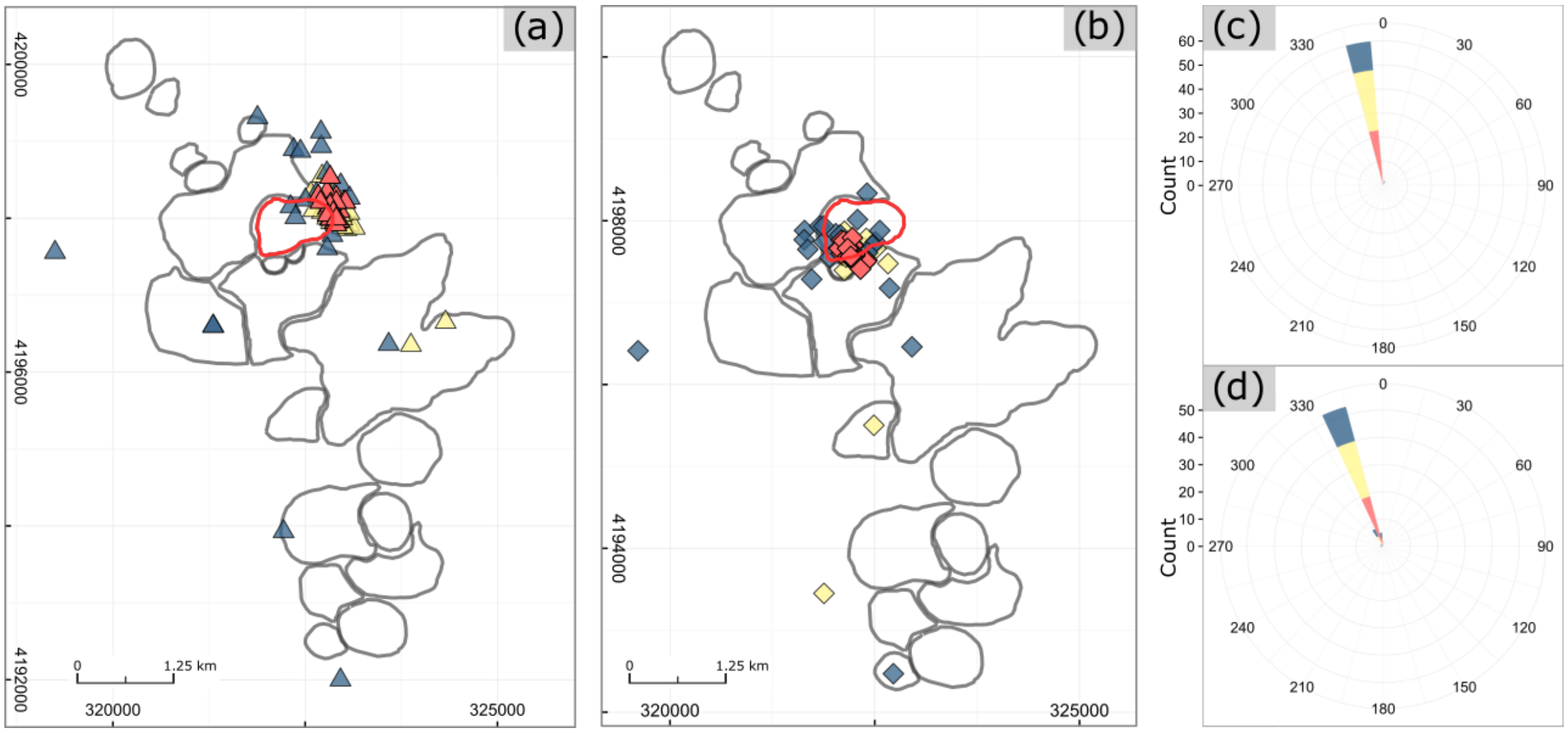

Figure 6

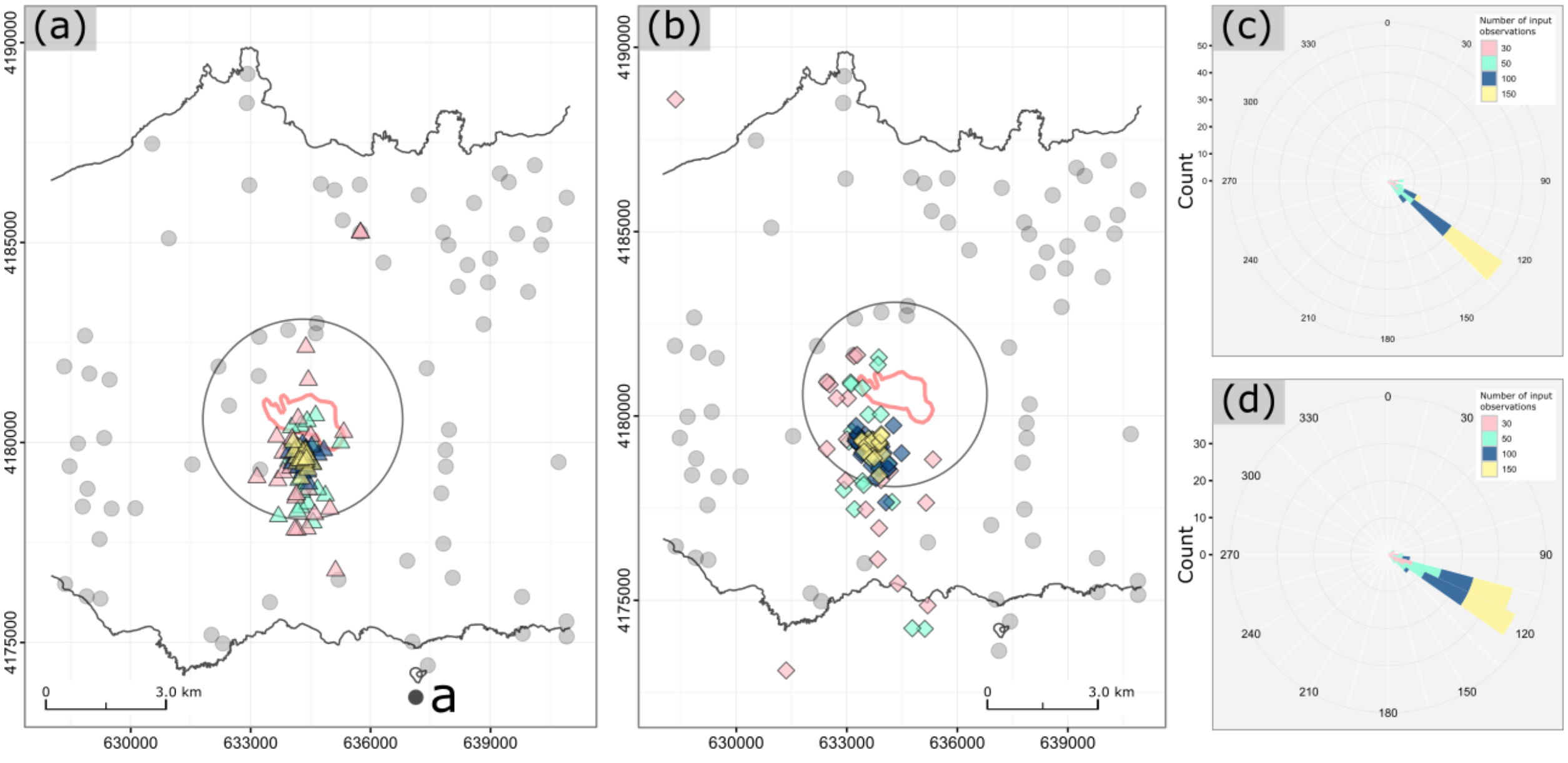

Figure 7

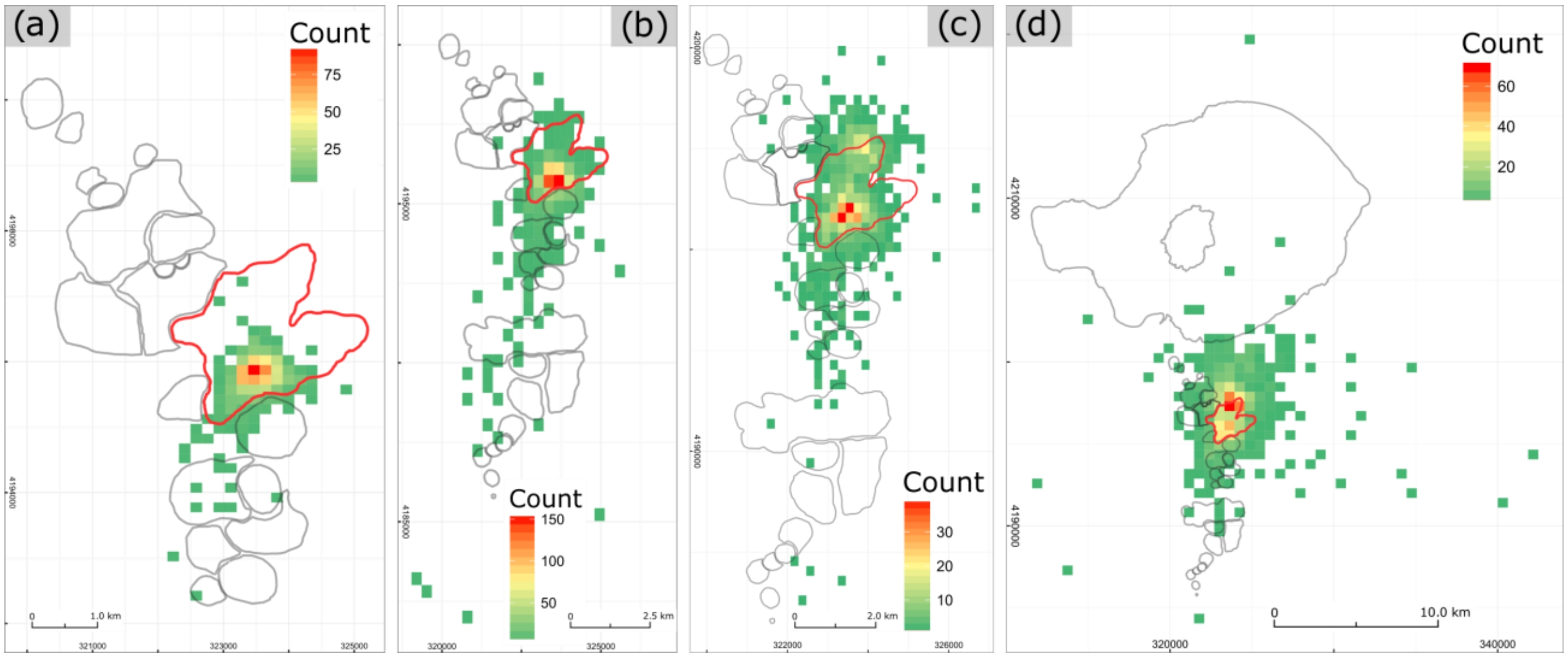

Figure 8

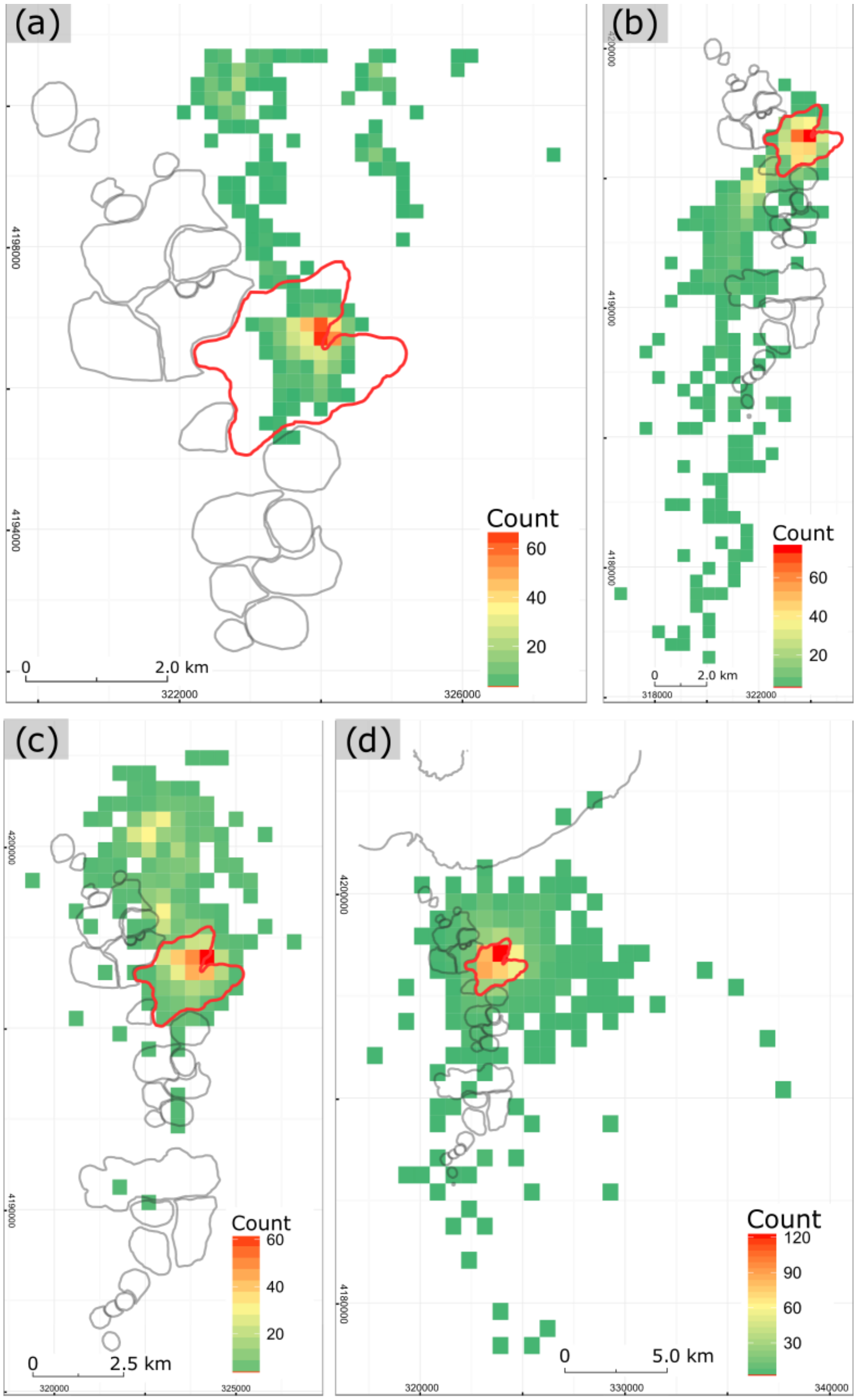

Figure 9

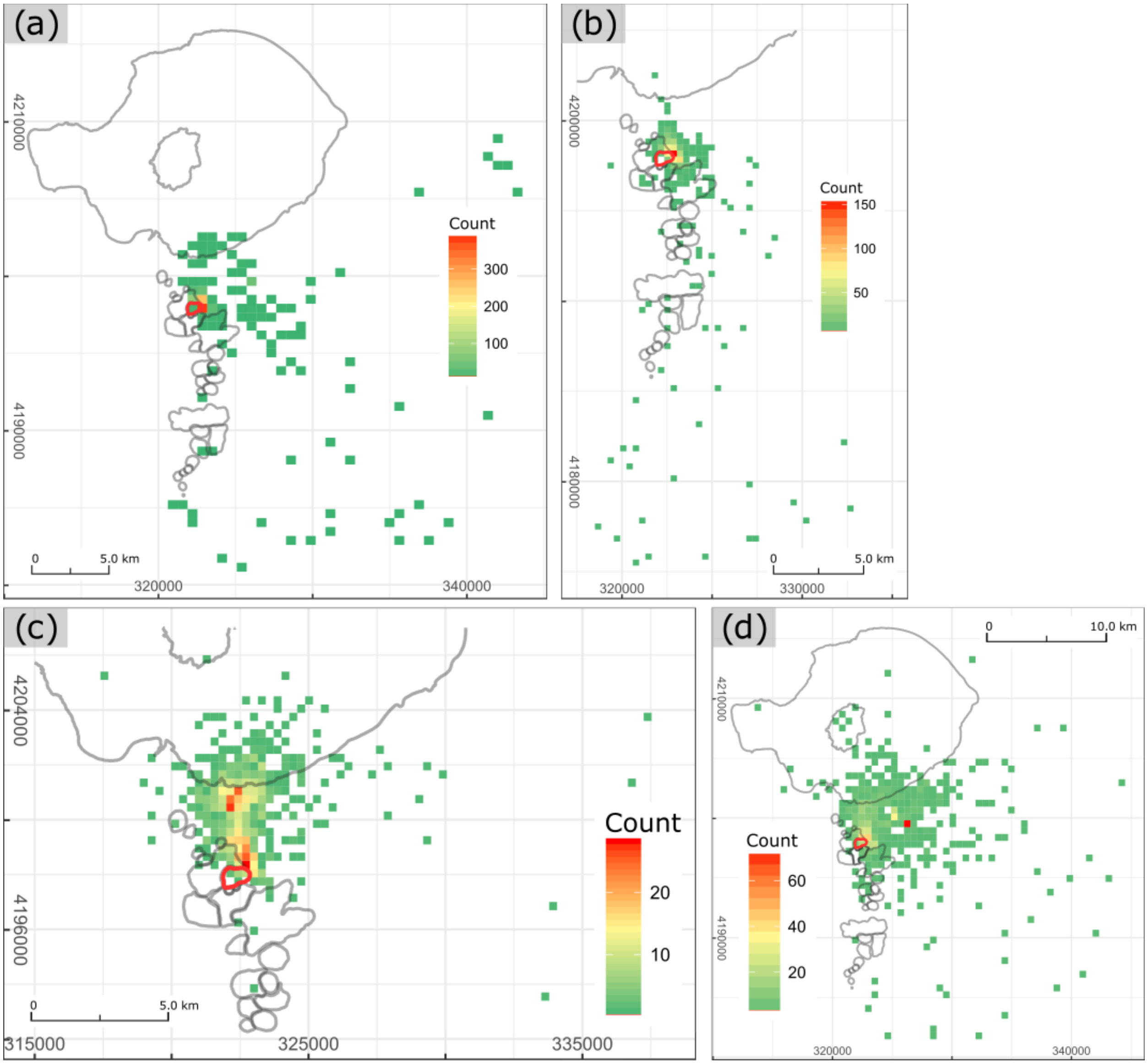

Figure 10

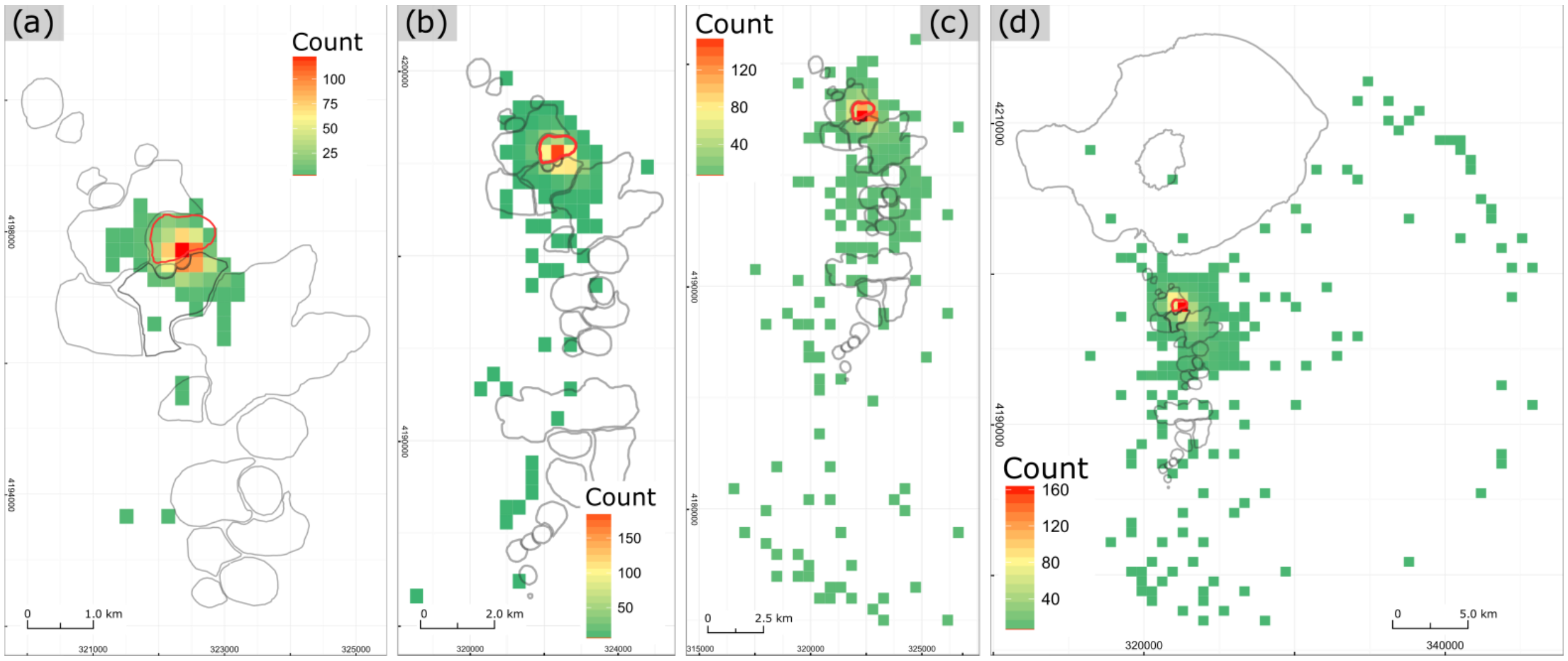

Figure 11

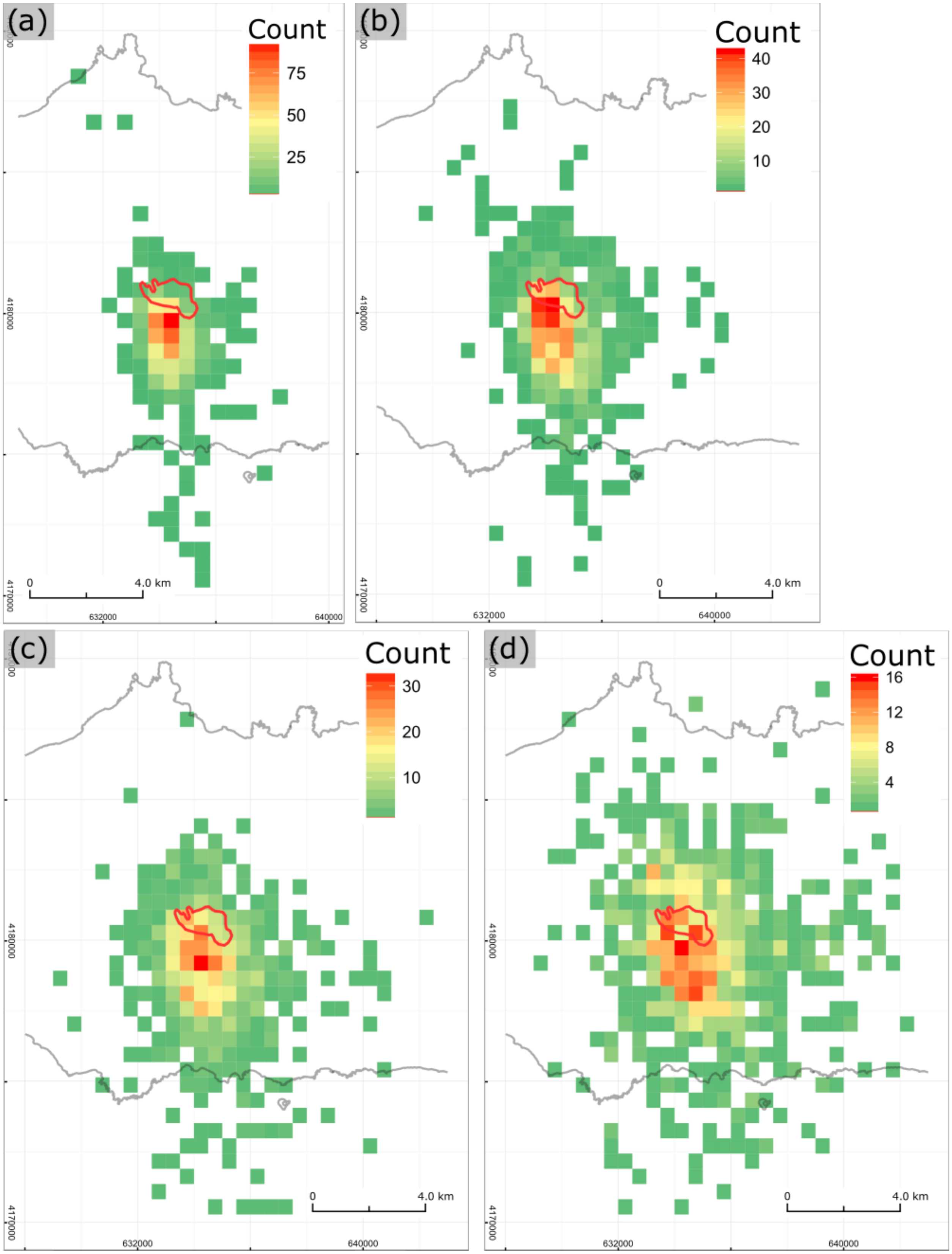

Figure 12

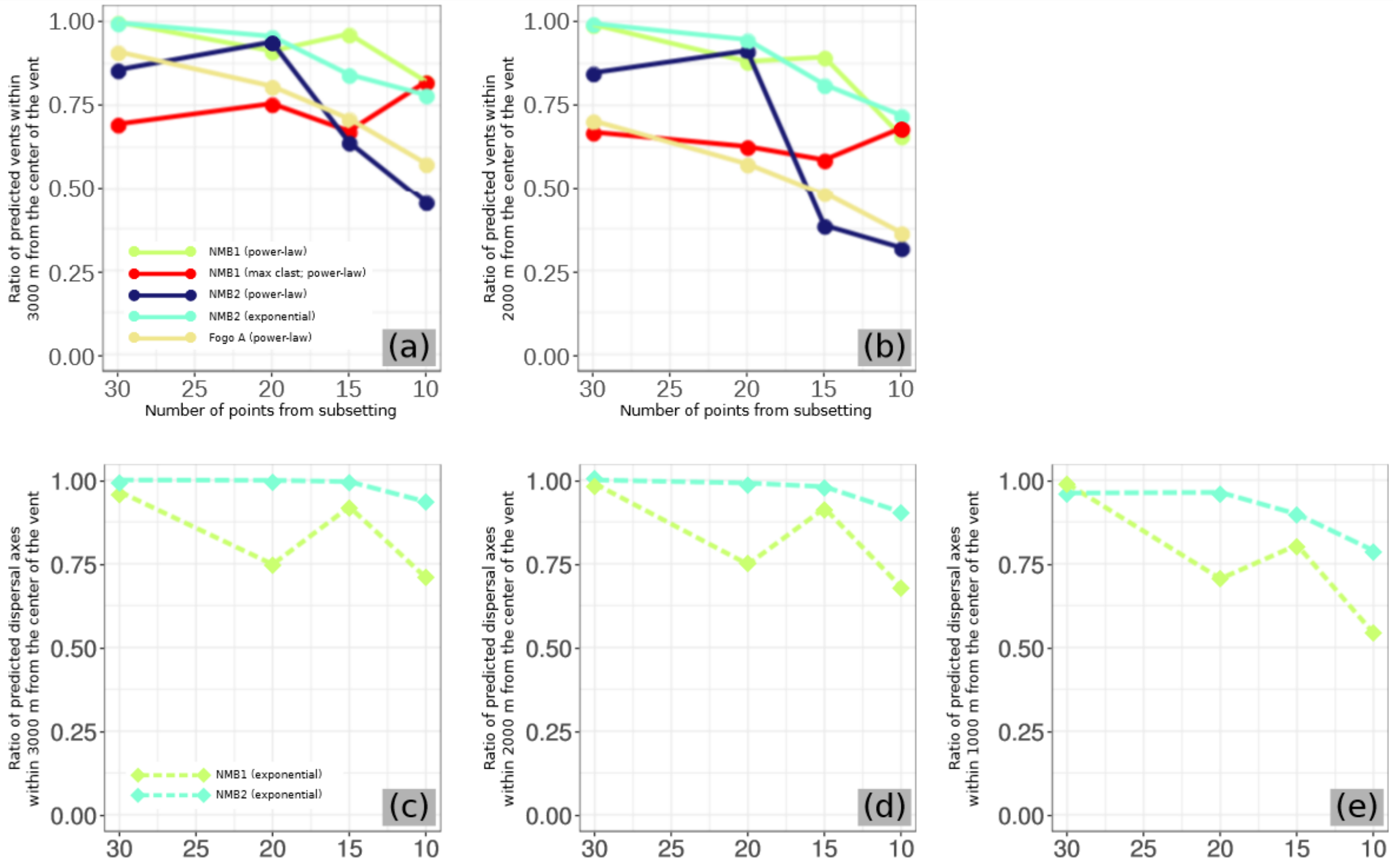

1.00
0.75
0.50
0.25
0.00
30
25
20
15
10
Ratio of predicted vents within 3000 m from the center of the vent
Number of points from subsetting
NMB1 (power-law)
NMB1 (max clast; power-law)
NMB2 (power-law)
NMB2 (exponential)
Fogo A (power-law)
(a)
Ratio of predicted vents within 2000 m from the center of the vent
(b)
Ratio of predicted dispersal axes within 3000 m from the center of the vent
NMB1 (exponential)
NMB2 (exponential)
(c)
Ratio of predicted dispersal axes within 2000 m from the center of the vent
(d)
Ratio of predicted dispersal axes within 1000 m from the center of the vent
(e)

Figure 13

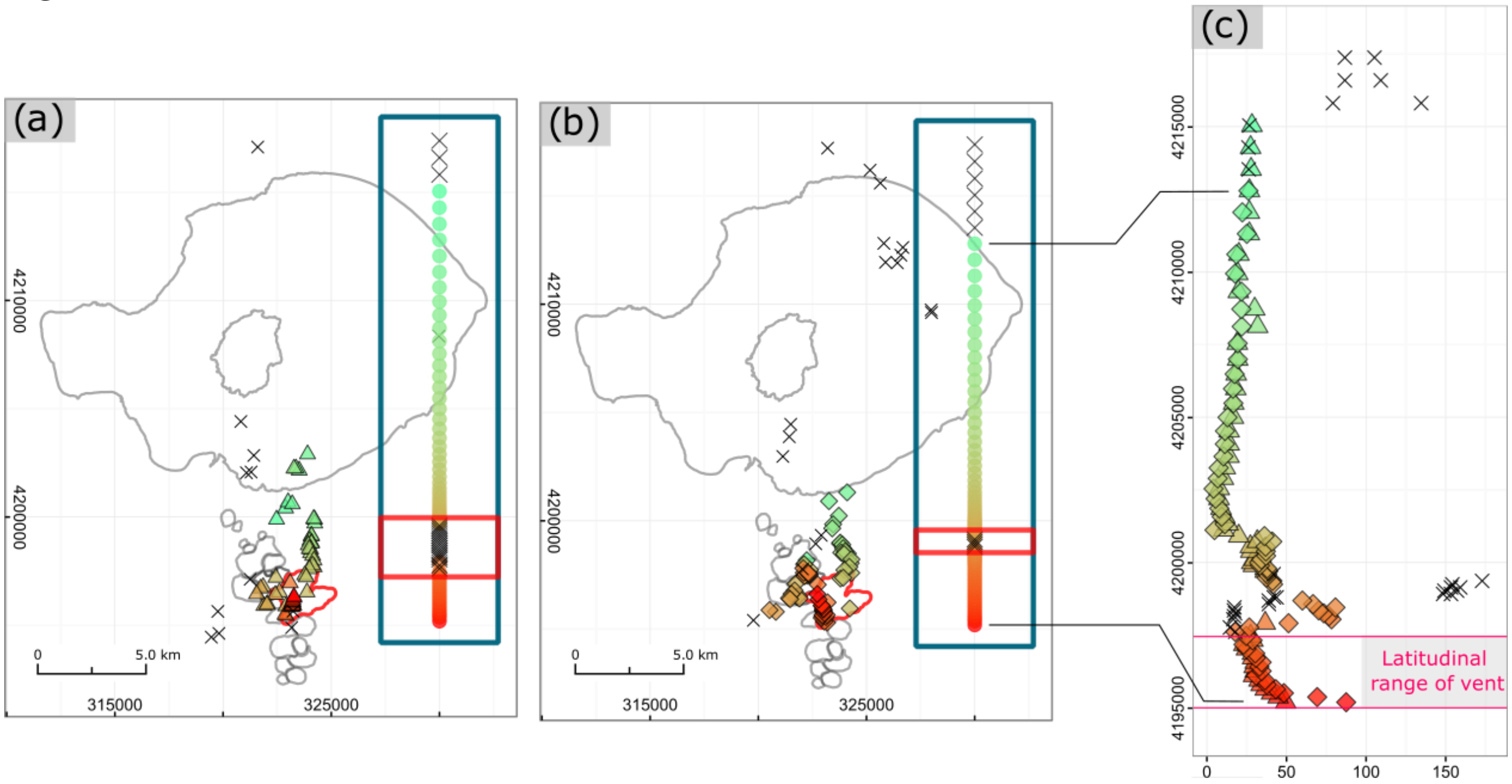

Figure 14

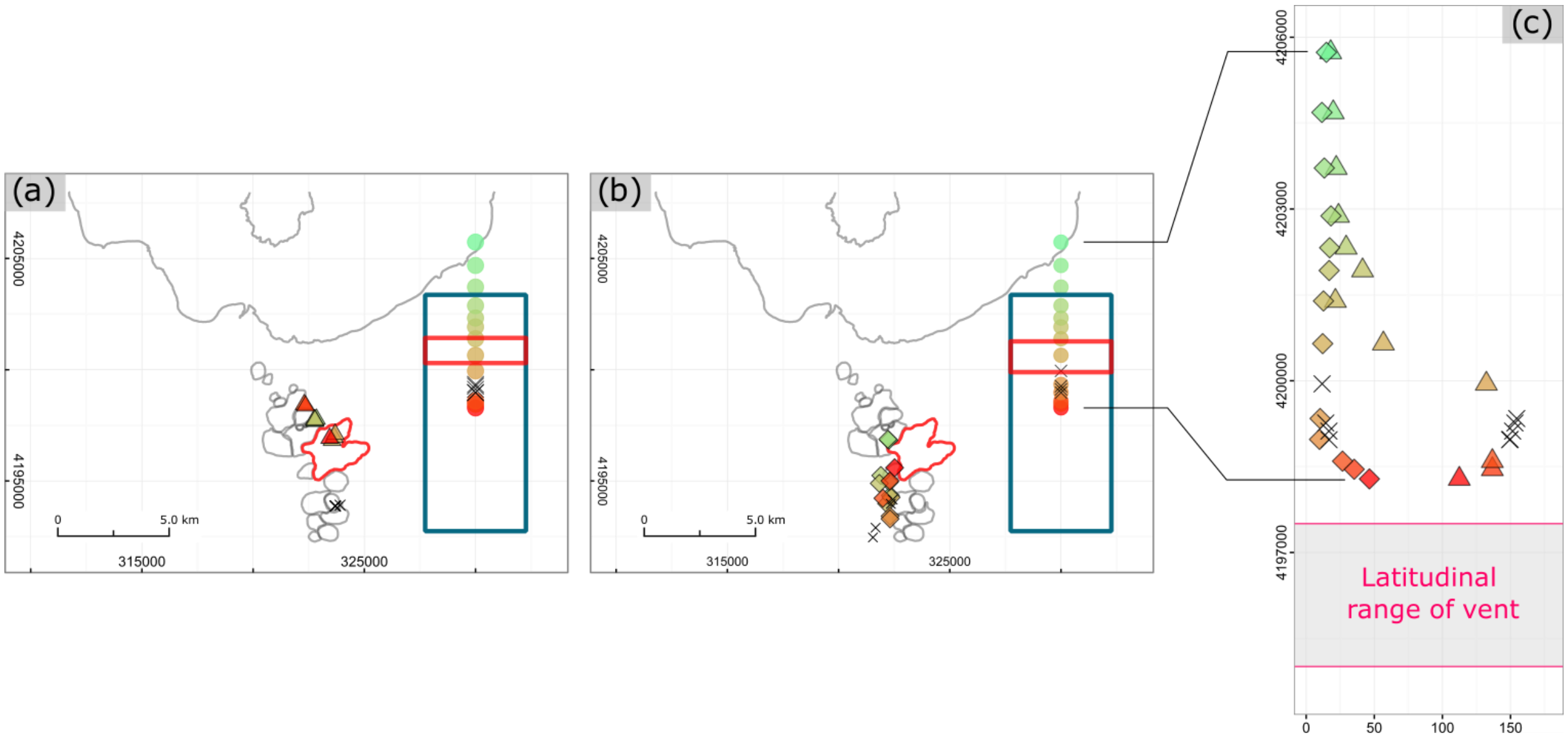

Figure 15

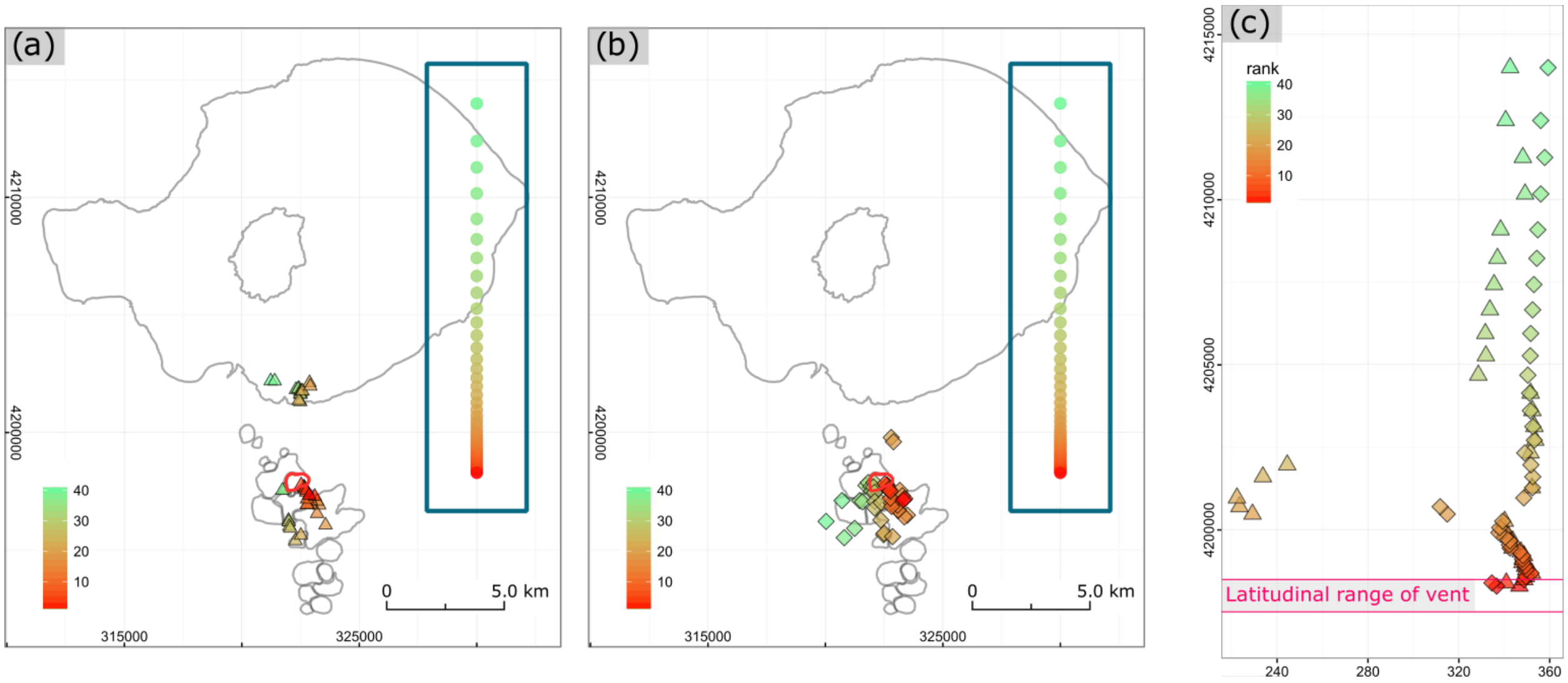

Figure 16

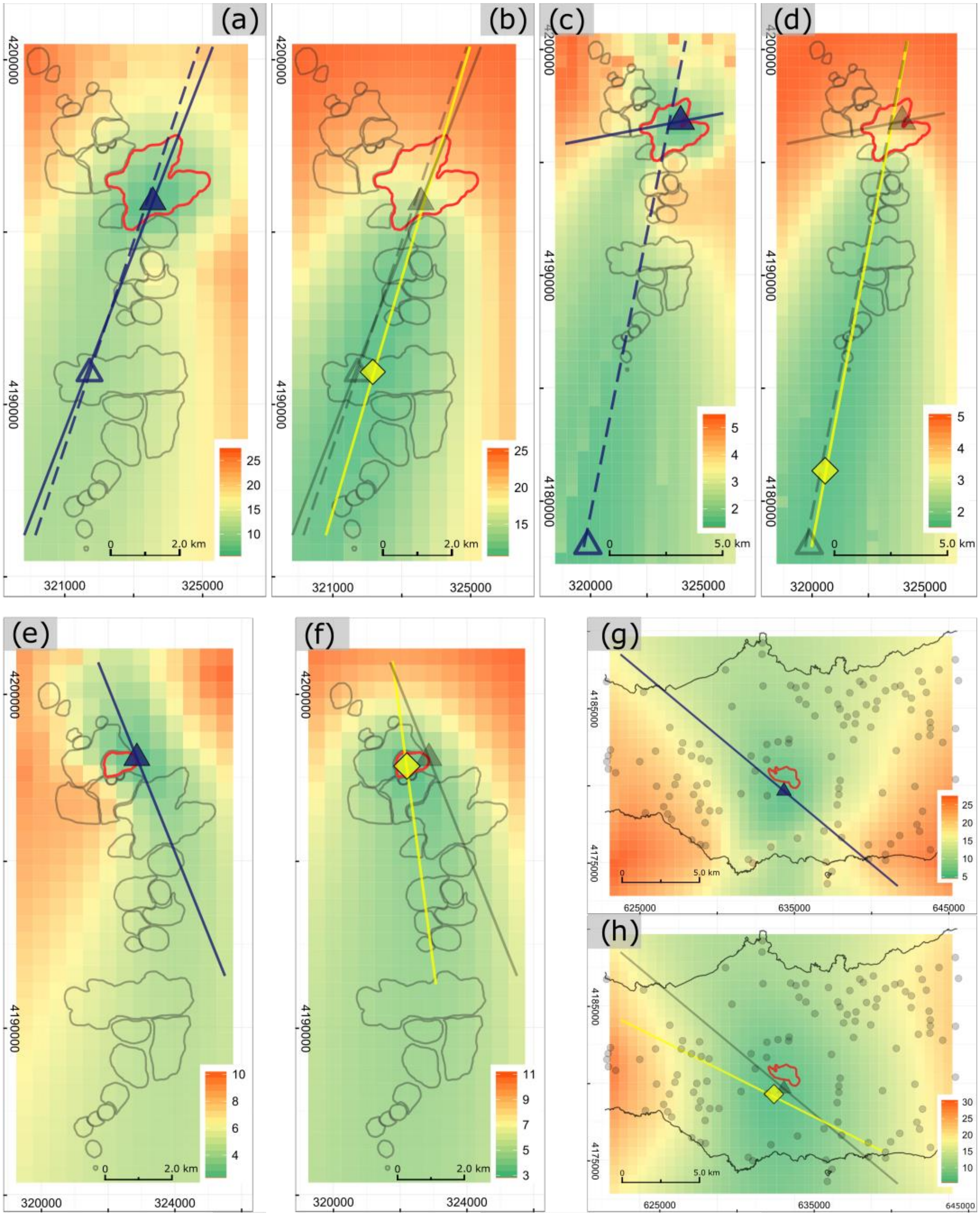

Figure 17

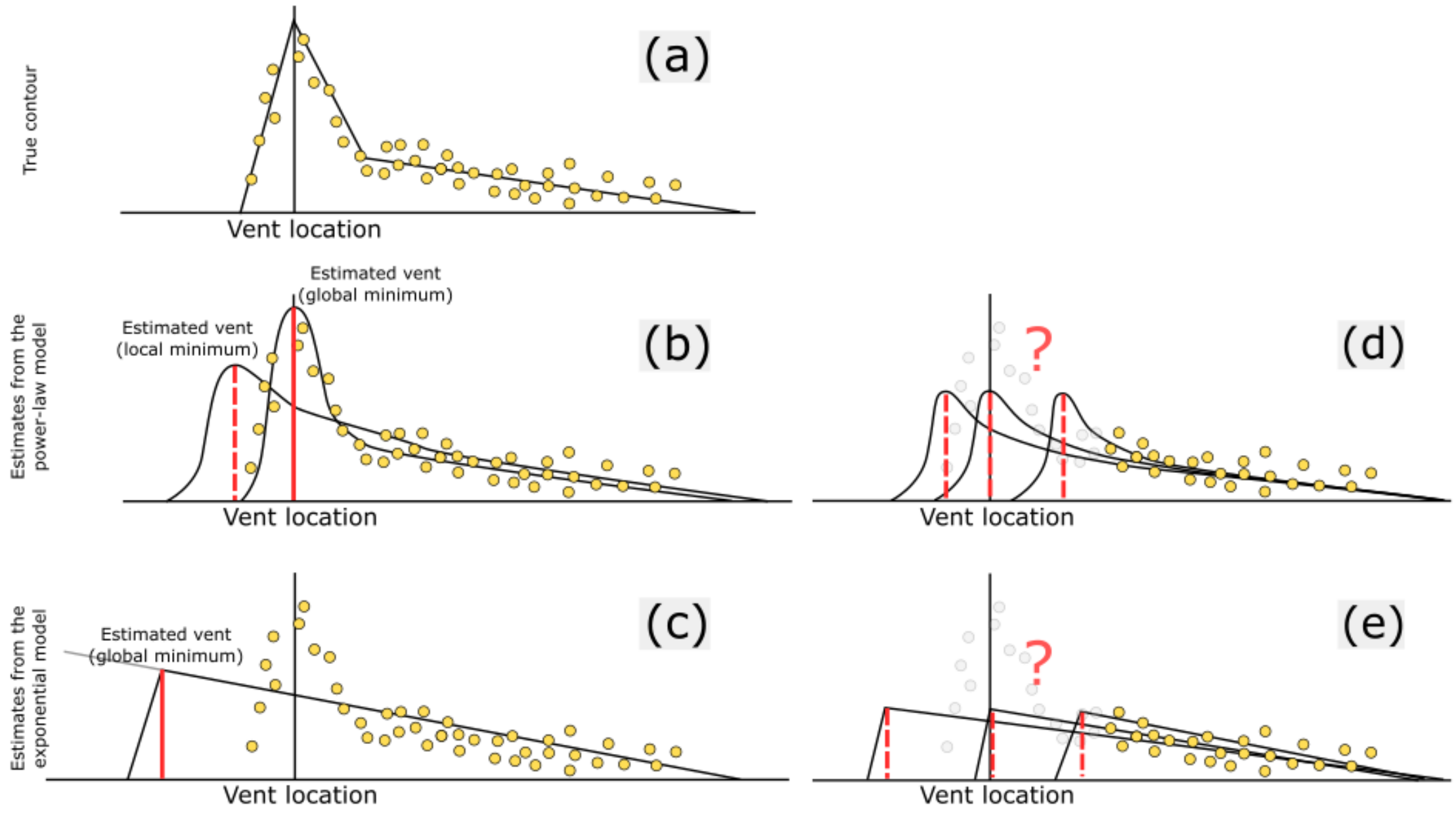

True contour
(a)
Vent location
Estimates from the power-law model
Estimated vent (global minimum)
Estimated vent (local minimum)
(b)
Vent location
(d)
?
Vent location
Estimates from the exponential model
Estimated vent (global minimum)
(c)
Vent location
(e)
?
Vent location

Figure 18

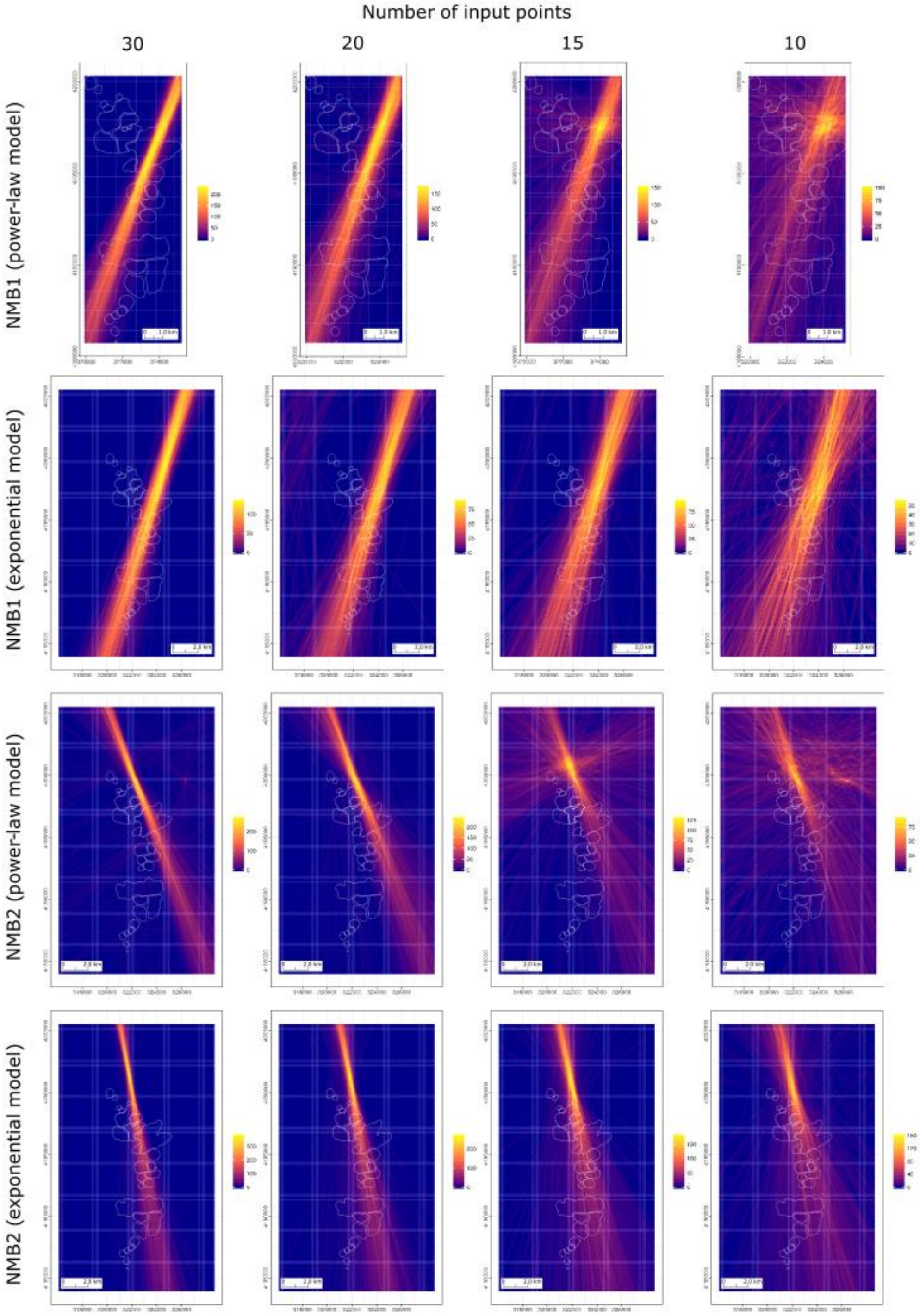

Figure 19

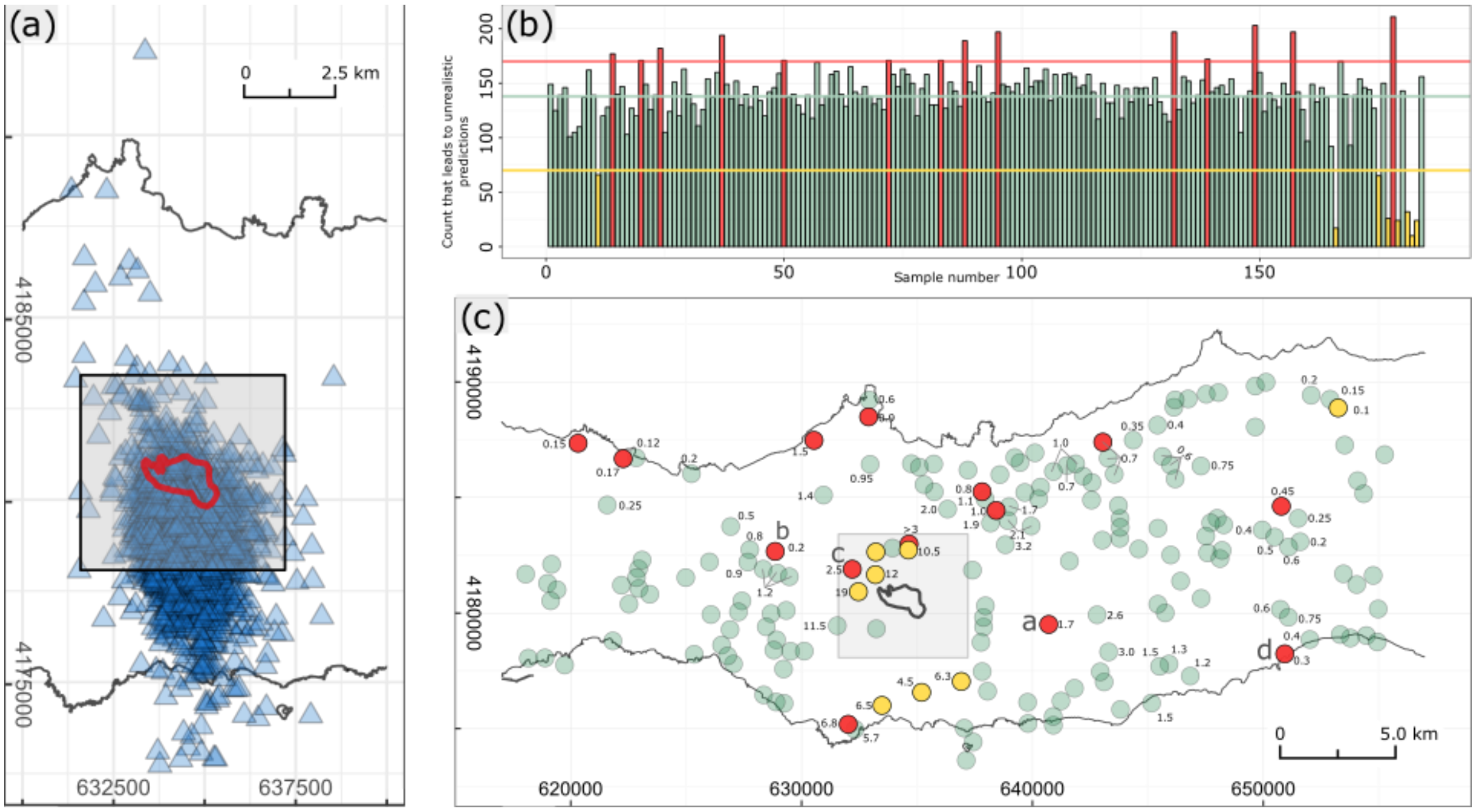

Figure 20

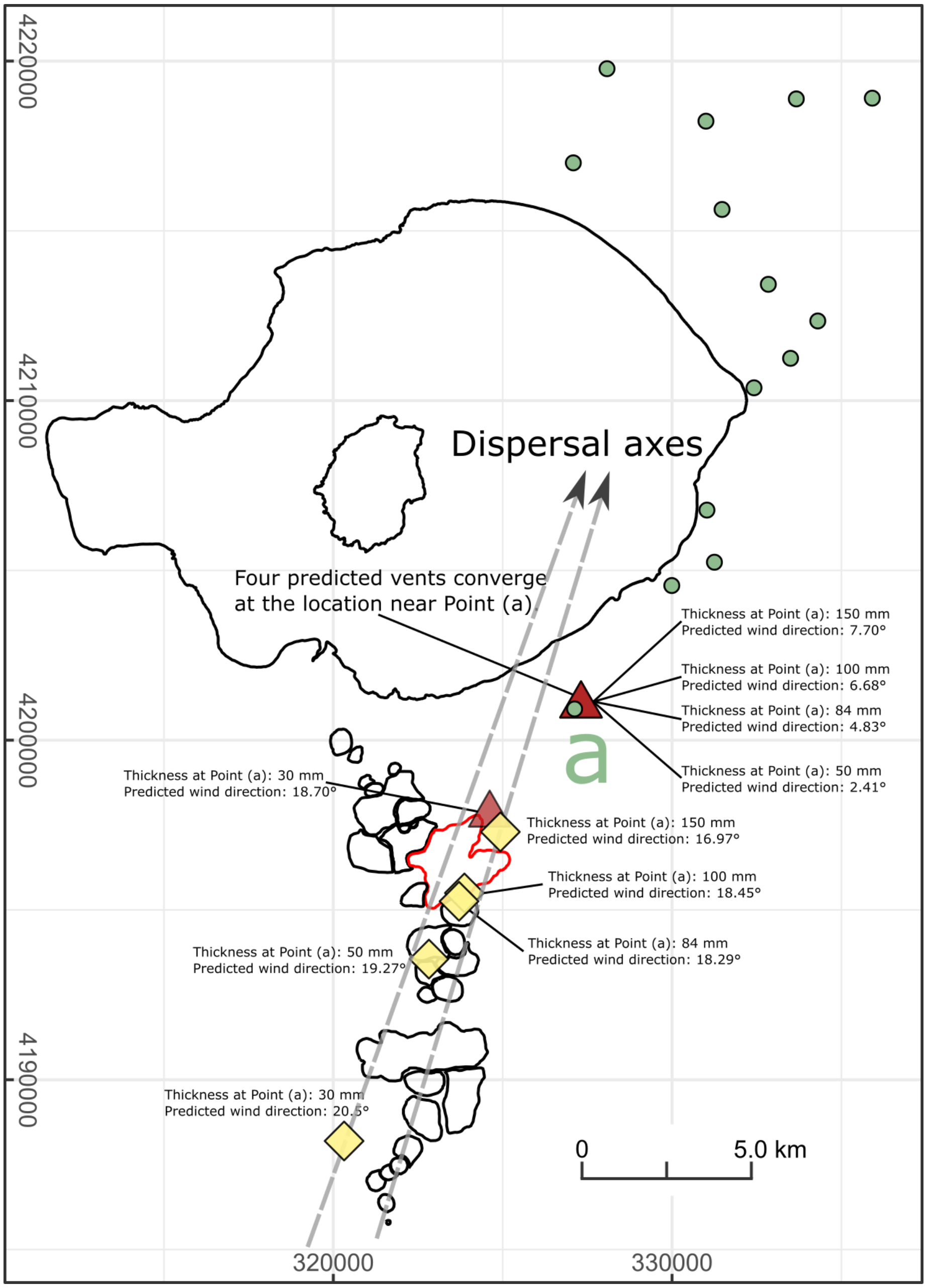

4220000
4210000
4200000
4190000
320000
330000
Dispersal axes
Four predicted vents converge at the location near Point (a)
Thickness at Point (a): 150 mm
Predicted wind direction: 7.70°
Thickness at Point (a): 100 mm
Predicted wind direction: 6.68°
Thickness at Point (a): 84 mm
Predicted wind direction: 4.83°
Thickness at Point (a): 50 mm
Predicted wind direction: 2.41°
a
Thickness at Point (a): 30 mm
Predicted wind direction: 18.70°
Thickness at Point (a): 150 mm
Predicted wind direction: 16.97°
Thickness at Point (a): 100 mm
Predicted wind direction: 18.45°
Thickness at Point (a): 84 mm
Predicted wind direction: 18.29°
Thickness at Point (a): 50 mm
Predicted wind direction: 19.27°
Thickness at Point (a): 30 mm
Predicted wind direction: 20.5°
0
5.0 km

Figure 21

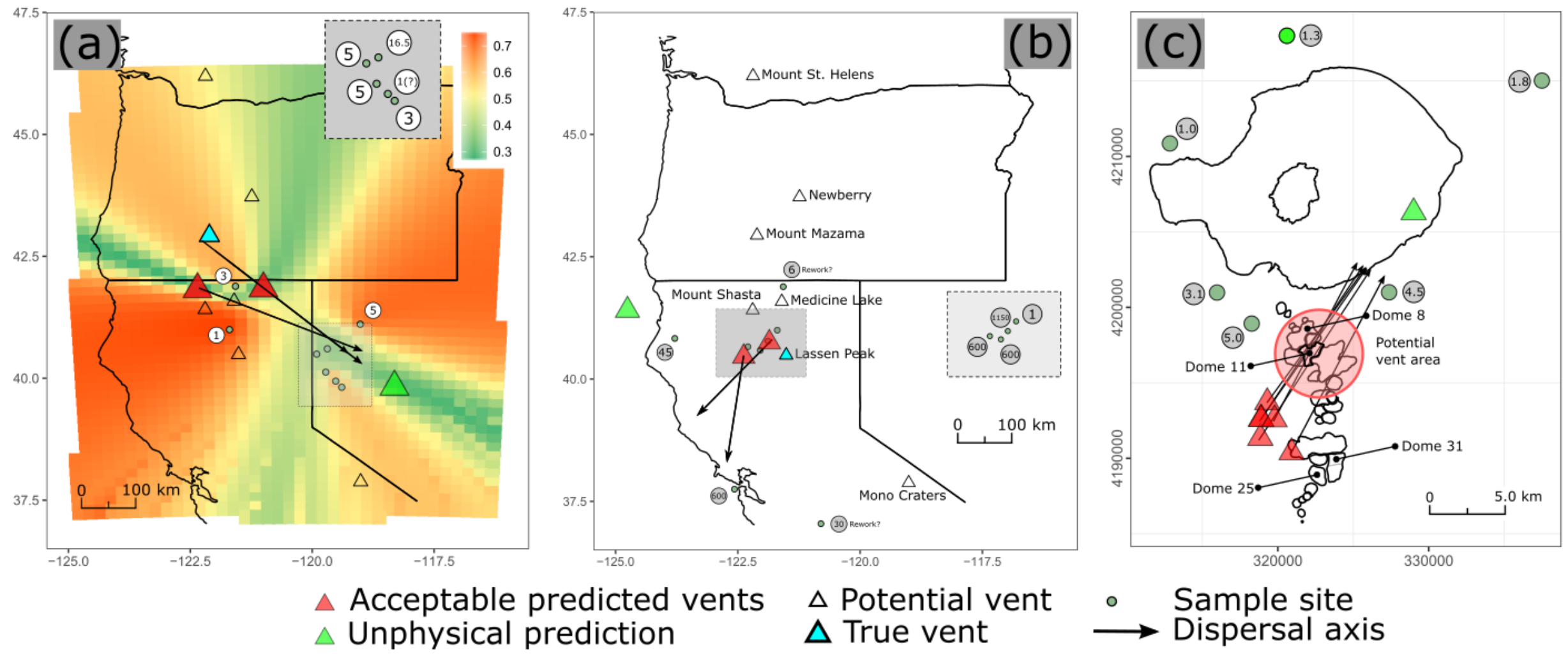